\documentclass[dvips]{siamltex}

\usepackage{amsfonts}
\usepackage{amsmath}
\usepackage{amssymb}
\usepackage{mathrsfs}
\usepackage{graphicx}
\usepackage[cspex,bbgreekl]{mathbbol}

\newcommand{\EE}{{\mathbb E }}  
\newcommand{\PP}{{\mathbb P }}
\newcommand{\II}{{\mathbb 1 }} 

\newcommand{\VV}{{\mathcal V }}

\newcommand{\SSS}{{\mathcal  S}}
\newcommand{\CC}{{\mathbb C}}

\newcommand{\tr}[1]{\mathrm{Tr}\left(#1\right)}

\title{Stabilizing feedback controls for quantum systems\thanks{This
work was supported by the ARO under Grant DAAD19-03-1-0073.}}

\author{
    Mazyar Mirrahimi\thanks{Centre Automatique et Syst\`emes,
    Ecole des Mines de Paris, 60 bd Saint-Michel, 75272 Paris Cedex
    06, France ({\tt mazyar.mirrahimi@polytechnique.org}).}
    \and Ramon Van Handel\thanks{Department of Physics
    and Control \& Dynamical Systems, California Institute of Technology
    266-33, Pasadena, CA 91125 USA ({\tt ramon@its.caltech.edu}).}
}

\begin{document}
\bibliographystyle{plain}
\maketitle

\begin{abstract}
    No quantum measurement can give full information on the state of a
    quantum system; hence any quantum feedback control problem is
    neccessarily one with partial observations, and can generally be
    converted into a completely observed control problem for an
    appropriate quantum filter as in classical stochastic control
    theory.  Here we study the properties of controlled quantum
    filtering equations as classical stochastic differential
    equations.  We then develop methods, using a combination of
    geometric control and classical probabilistic techniques, for
    global feedback stabilization of a class of quantum filters around
    a particular eigenstate of the measurement operator.
\end{abstract}



\section{Introduction}
\label{intro:sec}

Though they are both probabilistic theories, probability theory and
quantum mechanics have historically developed along very different lines.
Nonetheless the two theories are remarkably close, and indeed a rigorous
development of quantum probability \cite{maassen-qprob} contains classical
probability theory as a special case.  The embedding of classical into
quantum probability has a natural interpretation that is central to the
idea of a quantum measurement: any set of {\it commuting} quantum
observables can be represented as random variables on some probability
space, and conversely any set of random variables can be encoded as
commuting observables in a quantum model.  The quantum probability model
then describes the statistics of any set of measurements that we are
allowed to make, whereas the sets of random variables obtained from
commuting observables describe measurements that can be performed in a
single realization of an experiment.  As we are not allowed to make
noncommuting observations in a single realization, any quantum measurement
yields even in principle only partial information about the system.

\begin{figure}
\centering
\includegraphics[width=0.72\textwidth]{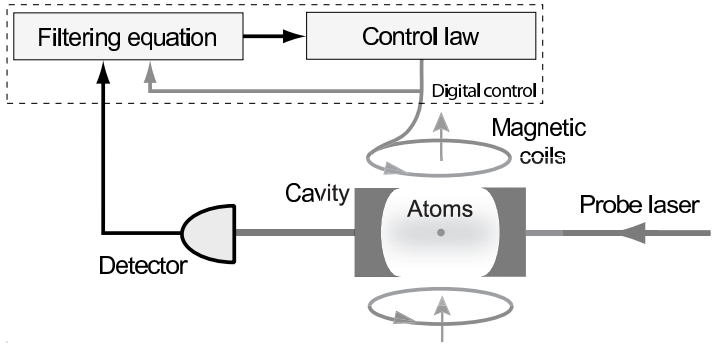}
\caption{A typical feedback control scenario in quantum optics.  A probe
laser scatters off a cloud of atoms in an optical cavity, and is
ultimately detected.  The detected signal is processed by a controller
which feeds back to the system through a time varying magnetic field.
\label{fig:model}}
\end{figure}

The situation in quantum feedback control
\cite{vanhandel-05,vanhandel-review} is thus very close to classical
stochastic control with partial observations \cite{Bensoussan1992}.  A
typical quantum control scenario, representative of experiments in quantum
optics, is shown in Fig.\ \ref{fig:model}.  We wish to control the state
of a cloud of atoms, e.g.\ we could be interested in controlling their
collective angular momentum. To observe the atoms, we scatter a laser
probe field off the atoms and measure the scattered light using a homodyne
detector (a cavity can be used to increase the interaction strength
between the light and the atoms).  The observation process is fed into a
controller which can feed back a control signal to the atoms through some
actuator, e.g.\ a time-varying magnetic field.  The entire setup can be
described by a Schr{\"o}dinger equation for the atoms and the probe field,
which takes the form of a ``quantum stochastic differential equation'' in
a Markovian limit.  The controller, however, only has access to the
observations of the probe.  The laser probe itself contributes quantum
fluctuations to the observations, hence the observation process can be
considered as a noisy observation of an atomic variable.

As in classical stochastic control we can use the properties of the
conditional expectation to convert the output feedback control problem
into one with complete observations.  The conditional expectation
$\pi_t(X)$ of an observable $X$ given the observations $\{Y_s:0\le s\le
t\}$ is the least mean square estimate of $X_t$ (the observable $X$ at
time $t$) given $Y_{s\le t}$.  One can obtain a quantum filtering equation
\cite{belavkin,bouten,BvHJ-05} that propagates $\pi_t(X)$, or
alternatively the conditional density matrix $\rho_t$ defined by the
relation $\pi_t(X)={\rm Tr}[\rho_t X]$.  This is the quantum counterpart
of the classical Kushner-Stratonovich equation, due to Belavkin
\cite{belavkin}, and plays an equivalent role in quantum stochastic
control. In particular, as $\mathbb{E}X_t=\mathbb{E}\pi_t(X)$ we can
control the expectations of observables by designing a state feedback
control law based on the filter.

Note that as the observation process $Y_{s\le t}$ is measured in a
single experimental realization, it is equivalent to a classical
stochastic process (i.e.\ the observables $Y_t$ commute with each
other at different times). But as the filter depends only on the
observations, it is thus equivalent to a classical stochastic
equation; in fact, the filter can be expressed as a classical
(It\^o) stochastic differential equation for the conditional
density matrix $\rho_t$.  Hence ultimately any quantum control
problem of this form is reduced to a classical stochastic control
problem for the filter.

In this paper we consider a class of quantum control problems of the
following form.  Rather than specifying a cost function to minimize, as in
optimal control theory, we desire to asymptotically prepare a particular
quantum state $\rho_f$ in the sense that $\mathbb{E}X_t\to{\rm Tr}[\rho_f
X]$ as $t\to\infty$ for all $X$ (for a deterministic version see
e.g.~\cite{mirrahimi-et-al2-04}). As $\mathbb{E}X_t=\mathbb{E}\pi_t(X)$,
this comes down to finding a feedback control that will ensure the
convergence $\rho_t\to\rho_f$ of the conditional density $\rho_t$. In
addition to this convergence, we will show that our controllers also
render the filter stochastically stable around the target state, which
suggests some degree of robustness to perturbations. In \S\ref{single:sec}
we will discuss the preparation of states in a cloud of atoms where the
$z$-component of the angular momentum has zero variance, whereas in
\S\ref{multi:sec} we will discuss the preparation of correlated states of
two spins.  Despite their relatively simple description the creation of
such states is not simple. Quantum feedback control may provide a
desirable method to reliably prepare such states in practice (though other
issues, e.g.\ the reduction of quantum filters \cite{vanhandel-05b} for
efficient real-time implementation, must be resolved before such schemes
can be realized experimentally; we refer to \cite{Geremia-science} for a
state-of-the-art experimental demonstration of a related quantum control
scenario.)

Though we have attempted to indicate the origin of the control problems
studied here, a detailed treatment of either the physical or mathematical
considerations behind our models is beyond the scope of this paper; for a
rigorous introduction to quantum probability and filtering we refer to
\cite{BvHJ-05}.  Instead we will consider the quantum filtering equation
as our starting point, and investigate the classical stochastic control
problem of feedback stabilization of this equation.  In \S\ref{geom:sec}
we first introduce some tools from stochastic stability theory and
stochastic analysis that we will use in our proofs.  In \S\ref{model:sec}
we introduce the quantum filtering equation and study issues such as
existence and uniqueness of solutions, continuity of the paths, etc.  In
\S\ref{single:sec} we pose the problem of stabilizing an angular momentum
eigenstate and prove global stability under a particular control law.
It is our expectation that the methods of \S\ref{single:sec} are
sufficiently flexible to be applied to a wide class of quantum state
preparation scenarios.  As an example, we use in \S\ref{multi:sec} the
techniques developed in \S\ref{single:sec} to stabilize particular
entangled states of two spins.
Additional results and numerical simulations will appear in 
\cite{MvHMM-05}.

\section{Geometric tools for stochastic processes}
\label{geom:sec}

In this section we briefly review two methods that will allow us to apply
geometric control techniques to stochastic systems.  The first is a
stochastic version of the classical Lyapunov and LaSalle invariance
theorems.  The second, a support theorem for stochastic differential
equations, will allow us to infer properties of stochastic sample paths
through the study of a related deterministic system.  We refer to the
references for proofs of the theorems.

\subsection{Lyapunov and LaSalle invariance theorems}
\label{lasalle:sec}

The Lyapunov stability theory and LaSalle's invariance theorem are
important tools in the analysis of and control design for deterministic
systems.  Similarly, their stochastic counterparts will play an essential
role in what follows.  The subject of stochastic stability was studied
extensively by Has'minski\u{\i} \cite{hasminskii} and by Kushner
\cite{kushner-67}.  We will cite a small selection of the results that
will be needed in the following: a Lyapunov (local) stability theorem for
Markov processes, and the LaSalle invariance theorem of Kushner
\cite{kushner-67,kushner-68,kushner-72}.

\begin{definition}
    Let $x_t^z$ be a diffusion process on the metric state space $X$,
    started at $x_0=z$, and let $\tilde z$ denote an equilibrium
    position of the diffusion, i.e.\ $x_t^{\tilde z}=\tilde z$.
    Then
    \begin{enumerate}
    \item the equilibrium $\tilde z$ is said to be {\rm stable in probability}
    if
    \begin{equation}
        \lim_{z\to\tilde z}\mathbb{P}\left(
            \sup_{0\le t<\infty}
            \|x_t^z-\tilde z\|\ge\varepsilon\right)
        =0\qquad \forall\varepsilon>0.
    \end{equation}
    \item the equilibrium $\tilde z$ is {\rm globally stable} if it is stable
    in probability and additionally
    \begin{equation}
        \mathbb{P}\left(\lim_{t\to\infty}x_t^z=\tilde z\right)=1\qquad
        \forall z\in X.
    \end{equation}
    \end{enumerate}
\end{definition}

In the following theorems we will make the following assumptions.
\begin{enumerate}
\item The state space $X$ is a complete separable metric space and $x_t^z$
    is a homogeneous strong Markov process on $X$ with continuous
    sample paths.
\item $V(\cdot)$ is a nonnegative real-valued continuous function on $X$.
\item For $\lambda>0$, let $Q_\lambda=\{x\in X:V(x)<\lambda\}$
    and assume $Q_\lambda$ is nonempty.  Let $\tau_\lambda=
    \inf\{t:x_t^z\not\in Q_\lambda\}$ and define the stopped process
    $\tilde x_t^z=x^z_{t\wedge\tau_\lambda}$.
\item   $\mathscr{A}_\lambda$ is the weak infinitesimal operator
    of $\tilde x_t$ and $V$ is in the domain of $\mathscr{A}_\lambda$.
\end{enumerate}
The following theorems can be found in Kushner \cite{kushner-67,kushner-68,kushner-72}.

\begin{theorem}[Local stability]
\label{thm:localstab}
    Let $\mathscr{A}_\lambda V\le 0$ in $Q_\lambda$.  Then the following
    hold:
    \begin{enumerate}
    \item $\lim_{t\to\infty}V(\tilde x_t^z)$ exists a.s., so $V(x_t^z)$
    converges for a.e.\ path remaining in $Q_\lambda$.
    \item $\mathbb{P}\mbox{\rm-lim}_{t\to\infty}\mathscr{A}_\lambda V(\tilde x_t^z)=0$,
    so $\mathscr{A}_\lambda V(x_t^z)\to 0$ in probability as $t\to\infty$
    for almost all paths which never leave $Q_\lambda$.
    \item For $z\in Q_\lambda$ and $\alpha\le\lambda$ we have the
    uniform estimate
    \begin{equation}
        \mathbb{P}\left(
            \sup_{0\le t<\infty}V(x_t^z)\ge\alpha
        \right)=
        \mathbb{P}\left(
            \sup_{0\le t<\infty}V(\tilde x_t^z)\ge\alpha
        \right)\le
        \frac{V(z)}{\alpha}.
    \end{equation}
    \item If $V(\tilde z)=0$ and $V(x)\ne 0$ for $x\ne\tilde z$,
    then $\tilde z$ in stable in probability.
    \end{enumerate}
\end{theorem}

The following theorem is a stochastic version of the LaSalle invariance
theorem.  Recall that a diffusion $x_t^z$ is said to be Feller continuous if
for fixed $t$, $\mathbb{E}G(x_t^z)$ is continuous in $z$ for any bounded
continuous $G$.

\begin{theorem}[Invariance]
\label{thm:lasalle}
    Let $\mathscr{A}_\lambda V\le 0$ in $Q_\lambda$.  Suppose $Q_\lambda$
    has compact closure, $\tilde x_t^z$ is Feller continuous, and
    that $\mathbb{P}(\|\tilde x_t^z-z\|>\varepsilon)
    \to 0$ as $t\to 0$ for any $\varepsilon>0$, uniformly for
    $z\in Q_\lambda$.  Then $\tilde x_t^z$ converges in probability to
    the largest invariant set contained in $C_\lambda=\{x\in Q_\lambda:
    \mathscr{A}_\lambda V(x)=0\}$.  Hence $x_t^z$ converges in
    probability to the largest invariant set contained in $C_\lambda$
    for almost all paths which never leave $Q_\lambda$.
\end{theorem}

\subsection{The support theorem}
\label{support:sec}

In the nonlinear control of deterministic systems an important role is
played by the application of geometric methods, e.g.\ Lie algebra
techniques, to the vector fields generating the control system.  Such
methods can usually not be directly applied to stochastic systems,
however, as the processes involved are not (sufficiently) differentiable.
The support theorem for stochastic differential equations, in its original
form due to Stroock and Varadhan \cite{stroock-varadhan}, connects events
of probability one for a stochastic differential equation to the solution
properties of an associated deterministic system.  One can then apply
classical techniques to the latter and invoke the support theorem to apply
the results to the stochastic system; see e.g.\ \cite{kunita-supp} for
the application of Lie algebraic methods to stochastic systems.

We quote the following form of the theorem \cite{kunita-flow,kunita-supp}.

\begin{theorem}
\label{thm:supportth}
    Let $M$ be a connected, paracompact $C^\infty$-manifold and let $X_k$,
    $k=0\ldots n$ be $C^\infty$ vector fields on $M$ such that all
    linear sums of $X_k$ are complete.
    Let $X_k=\sum_l X_k^l(x)\partial_l$ in local coordinates
    and consider the Stratonovich equation
    \begin{equation}
        dx_t=X_0(x_t)\,dt+\sum_{k=1}^n X_k(x_t)\circ dW_t^k,
        \qquad x_0=x.
    \end{equation}
    Consider in addition the associated deterministic control system
    \begin{equation}
        \frac{d}{dt}x_t^u=X_0(x_t^u)+\sum_{k=1}^n X_k(x_t^u)u^k(t),
        \qquad x_0^u=x
    \end{equation}
    with $u^k\in\mathscr{U}$, the set of all piecewise constant
    functions from $\mathbb{R}_+$ to $\mathbb{R}$.  Then
    \begin{equation}
        \mathscr{S}_x=
        \overline{\{x^u_\cdot:u\in\mathscr{U}^n\}}\subset
        \mathscr{W}_x
    \end{equation}
    where $\mathscr{W}_x$ is the set of all continuous paths from
    $\mathbb{R}_+$ to $M$ starting at $x$, equipped with the topology
    of uniform convergence on compact sets, and $\mathscr{S}_x$ is the
    smallest closed subset of $\mathscr{W}_x$ such that
    $\mathbb{P}(\{\omega\in\Omega:x_\cdot(\omega)\in\mathscr{S}_x\})=1$.
\end{theorem}

\section{Solution properties of quantum filters}
\label{model:sec}

The purpose of this section is to introduce the dynamical equations for
a general quantum system with feedback and to establish their basic
solution properties.

We will consider quantum systems with finite dimension $1<N<\infty$. The
state space of such a system is given by the set of density matrices
\begin{equation}
    \mathcal{S}=\{\rho\in\mathbb{C}^{N\times N}:
        \rho=\rho^*,~{\rm Tr}\,\rho=1,~\rho\ge 0\}
\end{equation}
where $\rho^*$ denotes Hermitian conjugation.  In noncommutative
probability the space $\mathcal{P}$ is the analog of the set of
probability measures of an $N$-state random variable. Finite-dimensional
quantum systems are ubiquitous in contemporary quantum physics; a
system with dimension $N=2^n$, for example, can represent the collective
state of $n$ qubits in the setting of quantum computing, and $N=2J+1$
represents a system with fixed angular momentum $J$.  The following lemma
describes the structure of $\mathcal{S}$:

\begin{lemma}
\label{l:hullc}
    $\mathcal{S}$ is the convex hull of
    $\{\rho\in\mathbb{C}^{N\times N}:
        \rho=vv^*,~v\in\mathbb{C}^N,~v^*v=1\}$.
\end{lemma}

\begin{proof}
    The statement is easily verified by diagonalizing the elements of
    $\mathcal{P}$.  \qquad
\end{proof}

We now consider continuous measurement of such a system, e.g.\ by weakly
coupling it to an optical probe field and performing a diffusive
observation of the field.  When the state of the system is conditioned on
the observation process we obtain the following matrix-valued It\^o
equation for the conditional density, which is a quantum analog of the
Kushner-Stratonovich equation of nonlinear filtering
\cite{belavkin,bouten,vanhandel-05}:
\begin{equation}
\label{eq:qfilt}
\begin{split}
    d\rho_t=-i(H_t\rho_t-\rho_t H_t)\,dt
     +(c\rho_t&c^* - \tfrac{1}{2}(c^*c\rho_t+\rho_tc^*c))\,dt \\
     &+\sqrt{\eta}\,(c\rho_t+\rho_tc^*
        -{\rm Tr}[(c+c^*)\rho_t]\rho_t)\,dW_t.
\end{split}
\end{equation}
Here we have introduced the following quantities:
\begin{itemize}
\item The Wiener process $W_t$ is the innovation
$dW_t=dy_t-\sqrt{\eta}\,{\rm Tr}[(c+c^*)\rho_t]dt$. Here $y_t$, a
continuous semimartingale with quadratic variation $\langle
y,y\rangle_t=t$, is the observation process obtained from the system.
\item $H_t=H_t^*$ is a Hamiltonian matrix which describes the action of
external forces on the system.  We will consider $H_t$ of the form
$H_t=F+u_tG$ with $F=F^*$, $G=G^*$ and the (real) scalar control input
$u_t$.
\item $u_t$ is a bounded real c{\`a}dl{\`a}g process that is adapted to
$\mathcal{F}_t^y=\sigma(y_s,0\le s\le t)$, the filtration generated by the
observations up to time $t$.
\item  $c$ is a matrix which determines the coupling to the external
(readout) field.
\item $0<\eta\le 1$ is the detector efficiency.
\end{itemize}
Let us begin by studying a different form of the equation (\ref{eq:qfilt}).
Consider the linear It\^o equation
\begin{equation}
\label{eq:zakai}
    d\tilde\rho_t=-i(H_t\tilde\rho_t-\tilde\rho_t H_t)\,dt
     +(c\tilde\rho_tc^* -
        \tfrac{1}{2}(c^*c\tilde\rho_t+\tilde\rho_tc^*c))\,dt
     +\sqrt{\eta}\,(c\tilde\rho_t+\tilde\rho_tc^*)\,dy_t,
\end{equation}
which is the quantum analog of the Zakai equation.  As it obeys a 
global (random) Lipschitz condition, this equation has a unique strong 
solution (\cite{protter}, pp.\ 249--253).

\begin{lemma}
\label{lem:invzakai}
    The set of nonnegative nonzero matrices is a.s.\ invariant
    for {\rm (\ref{eq:zakai})}.
\end{lemma}

\begin{proof}
    We begin by expanding $\tilde\rho_0$ into its eigenstates,
    i.e.\ $\tilde\rho_0=\sum_i\lambda_iv_0^{i}v_0^{i*}$ with
    $v_0^{i}\in\mathbb{C}^N$ being the $i$th eigenvector and $\lambda_i$
    the $i$th eigenvalue. As $\tilde\rho_0$ is nonnegative all the
    $\lambda_i$ are nonnegative.
    Now consider the set of equations
    \begin{equation}
    \label{eq:flibbered}
        d\rho_t^{i}=-i(H_t\rho_t^{i}
            -\rho_t^{i}H_t)\,dt
         +(c\rho_t^{i}c^* -
            \tfrac{1}{2}(c^*c\rho_t^{i}+\rho_t^{i}c^*c))\,dt
         +(c\rho_t^{i}+\rho_t^{i}c^*)\,dW_t'
    \end{equation}
    with $\rho_0^{i}=v_0^{i}v_0^{i*}$.  Here we have extended our
    probability space to admit a Wiener process $\hat W_t$ that is
    independent of $y_t$, and $W'_t=\sqrt{\eta}\,y_t+\sqrt{1-\eta}\,\hat W_t$.
    The process $\tilde\rho_t$ is then equivalent in law to
    $\mathbb{E}[\rho_t'|\mathcal{F}_t^y]$, where $\rho_t'=
    \sum_i\lambda_i\rho_t^{i}$.

    Now note that the solution of the set of equations
    \begin{equation}
    \label{eq:linlin}
        dv_t^{i}=-iH_tv_t^{i}\,dt
        -\tfrac{1}{2}c^*c\,v_t^{i}\,dt+
            c\,v_t^{i}\,dW_t',
        ~~~~~~~ ~~~~~~~ v_t^{i}\in\mathbb{C}^N
    \end{equation}
    satisfies $\rho_t^{i}=v_t^{i}v_t^{i*}$, as is readily verified by
    It\^o's rule.  By \cite{protter}, pp.\ 326 we have that
    $v_t^{i}=U_tv_0^{i}$ where the random matrix $U_t$ is a.s.\
    invertible for all $t$.  Hence a.s.\ $v_t^{i}\ne 0$ for any
    finite time unless $v_0^{i}=0$.  Thus clearly $\rho_t'$ is a.s.\ a
    nonnegative nonzero matrix for all $t$, and the
    result follows. \qquad
\end{proof}

\begin{proposition}
    Eq.\ {\rm (\ref{eq:qfilt})} has a unique strong solution
    $\rho_t=\tilde\rho_t/{\rm Tr}\,\tilde\rho_t$ in $\mathcal{S}$.
\end{proposition}

Clearly this must be satisfied if (\ref{eq:qfilt}) is to propagate a density.

\begin{proof}
    As the set of nonnegative nonzero matrices is invariant for
    $\tilde\rho_t$, this implies in particular that
    ${\rm Tr}\,\tilde\rho_t>0$ for all $t$ a.s.  Thus the result
    follows simply from application of It\^o's rule to (\ref{eq:zakai}),
    and from the fact that if $M=\sum_i\lambda_iv_i$ is a nonnegative
    nonzero matrix, then $M/{\rm Tr}\,M=
    \sum_i(\lambda_i/\sum_j\lambda_j)v_i\in\mathcal{S}$. \qquad
\end{proof}

\begin{proposition}
\label{pro:uniformstoch}
    The following uniform estimate holds for {\rm (\ref{eq:qfilt})}:
    \begin{equation}
        \mathbb{P}\left(\sup_{0\le\delta\le\Delta}
            \|\rho_{t+\delta}-\rho_t\|>\varepsilon\right)
        \le C\Delta(1+\Delta) \qquad \forall\varepsilon>0
    \end{equation}
    where $0<C<\infty$ depends only on $\varepsilon$ and
    $\|\cdot\|$ is the Frobenius norm.  Hence the solution of {\rm
    (\ref{eq:qfilt})} is stochastically continuous uniformly in $t$
    and $\rho_0$.
\end{proposition}

\begin{proof}
    Write $\rho_t=\rho_0+\Phi_t+\Xi_t$ where
    \begin{equation}
    \label{eq:uniformphit}
        \Phi_t=
        \int_0^t\left[-i(H_s\rho_s-\rho_s H_s)
        +(c\rho_sc^* - \tfrac{1}{2}(c^*c\rho_s+\rho_sc^*c))
        \right]ds,
    \end{equation}
    \begin{equation}\label{eq:uniformxi}
        \Xi_t=
        \int_0^t
         \sqrt{\eta}\,
        (c\rho_s+\rho_sc^*
        -{\rm Tr}[(c+c^*)\rho_s]\rho_s)\,dW_s.
    \end{equation}
    For $\Xi_t$ we have the estimate (\cite{arnold}, pp.\ 81)
    \begin{equation}
    \label{eq:uniformxit}
        \mathbb{E}\left(\sup_{0\le\delta\le\Delta}
            \|\Xi_{t+\delta}-\Xi_t\|^2\right)\le
        4\eta
        \int_t^{t+\Delta}\mathbb{E}\|
            c\rho_s+\rho_sc^*-{\rm Tr}[(c+c^*)\rho_s]\rho_s\|^2
        \,ds.
    \end{equation}
    As the integrand is bounded clearly this expression is bounded by
    $C_1\Delta$ for some positive constant $C_1<\infty$.
    For $\Phi_t$ we can write
    \begin{equation}
        \mathbb{E}\left(
        \sup_{0\le\delta\le\Delta}\|\Phi_{t+\delta}-\Phi_t\|^2\right)
        \le \mathbb{E}\left[\sup_{0\le\delta\le\Delta}\int_t^{t+\delta}
        \|G_s\|\,ds\right]^2
        =\mathbb{E}\left[\int_t^{t+\Delta}\|G_s\|\,ds\right]^2
    \end{equation}
    where $G_s$ denotes the integrand of (\ref{eq:uniformphit}).
    As $\|G_s\|$ is bounded we can estimate this expression by
    $C_2\Delta^2$ with $C_2<\infty$.
    Using $\|A+B\|^2\le 2(\|A\|^2+\|B\|^2)$ we can write
    \begin{equation}
        \sup_{0\le\delta\le\Delta}\|\rho_{t+\delta}-\rho_t\|^2
        \le 2\left(\sup_{0\le\delta\le\Delta}\|\Phi_{t+\delta}-\Phi_t\|^2+
        \sup_{0\le\delta\le\Delta}\|\Xi_{t+\delta}-\Xi_t\|^2\right).
    \end{equation}
    Finally, Chebychev's inequality gives
    \begin{equation}
        \mathbb{P}\left(\sup_{0\le\delta\le\Delta}
            \|\rho_{t+\delta}-\rho_t\|>\varepsilon\right)\le
        \frac{1}{\varepsilon^2}
        \mathbb{E}\left(\sup_{0\le\delta\le\Delta}
            \|\rho_{t+\delta}-\rho_t\|^2\right)
        \le \frac{2C_1\Delta+2C_2\Delta^2}{\varepsilon^2}
    \end{equation}
    from which the result follows. \qquad
\end{proof}

{\em Remark.} The statistics of the observation process $y_t$ should of
course depend both on the control $u_t$ that is applied to the system and
on the initial state $\rho_0$.  We will always assume that the filter
initial state $\rho_0$ matches the state in which the system is initially
prepared (i.e.\ we do not consider ``wrongly initialized'' filters) and
that the same control $u_t$ is applied to the system and to the filter
(see Fig.\ \ref{fig:model}).  Quantum filtering theory then guarantees
that the innovation $W_t$ is a Wiener process.  To simplify our proofs, we 
make from this point on the following choice: for all initial states and 
control policies, the corresponding observation processes are defined in 
such a way that they give rise to the same innovation process 
$W_t$\footnote{
This is quite contrary to the usual choice in stochastic control theory:
there the system and observation noise are chosen to be fixed Wiener
processes, and every initial state and control policy give rise to a
different innovation (Wiener) process.  However, in the quantum case the
system and observation noise do not even commute with the observations
process, and thus we cannot use them to fix the innovations.  In fact, the
observation process $y_t$ that emerges from the quantum probability model
is only defined in a ``weak'' sense as a $^*$-isomorphism between an
algebra of observables and a set of random variables on
$(\Omega,\mathcal{F},\mathbb{P})$ \cite{BvHJ-05}.  Hence we might as well
choose the isomorphism for each initial state and control in such a way
that all observations $y_t[\rho_0,u_t]$ give rise to the fixed innovations 
process $W_t$, regardless of $\rho_0,u_t$.  That such an isomorphism 
exists is evident from the form of the filtering equation at least in the 
case that $u_t$ is a functional of the innovations  (e.g.\ if 
$u_t=u(\rho_t)$): if we calculate the strong solution of (\ref{eq:qfilt}) 
given a fixed driving process $W_t$, $\rho_0$, and $u_t[W]$, then $dy_t = 
dW_t + \sqrt{\eta}\,{\rm Tr}[(c+c^*)\rho_t]dt$ must have the same law as 
$y_t[\rho_0,u_t]$.

Note that the only results that depend on the precise choice of
$y_t[\rho_0,u_t]$ on $(\Omega,\mathcal{F},\mathbb{P})$ are joint
statistics of the filter sample paths for different initial states or
controls.  However, such results are physically meaningless as the
corresponding quantum models generally do not commute.
}.

We now specialize to the following case:
\begin{itemize}
\item $u_t=u(\rho_t)$ with $u\in C^1(\mathcal{S},\mathbb{R})$.
\end{itemize}
In this simple feedback case we can prove several important properties of
the solutions.  First, however, we must show existence and uniqueness for 
the filtering equation with feedback: it is not a priori obvious that the 
feedback $u_t=u(\rho_t)$ results in a well-defined c{\`a}dl{\`a}g control.

\begin{proposition}
\label{pro:feedbackxu}
	Eq.\ {\rm (\ref{eq:qfilt})} with $u_t=u(\rho_t)$, $u\in C^1$ and
	$\rho_0=\rho\in\mathcal{S}$ has a unique strong solution 
	$\rho_t\equiv\varphi_t(\rho,u)$ in $\mathcal{S}$, and
	$u_t$ is a continuous bounded control.
\end{proposition}

\begin{proof}
As $\mathcal{S}$ is compact, we can find an open set $\mathcal{T}\subset
\mathbb{C}^{N\times N}$ such that $\mathcal{S}$ is strictly contained in
$\mathcal{T}$.  Let $C(\rho):\mathbb{C}^{N\times N}\to[0,1]$ be a smooth
function with compact support such that $C(\rho)=1$ for
$\rho\in\mathcal{T}$, and let $U(\rho)$ be a $C^1(\mathbb{C}^{N\times
N},\mathbb{R})$ function such that $U(\rho)=u(\rho)$ for
$\rho\in\mathcal{S}$.  Then the equation
\begin{multline*}
    d\bar\rho_t=-iC(\bar\rho_t)[F+U(\bar\rho_t)G,\bar\rho_t]\,dt
     +C(\bar\rho_t)(c\bar\rho_tc^* - \tfrac{1}{2}(c^*c\bar\rho_t+\bar\rho_tc^*c))\,dt \\
     +C(\bar\rho_t)\sqrt{\eta}\,(c\bar\rho_t+\bar\rho_tc^*
        -{\rm Tr}[(c+c^*)\bar\rho_t]\bar\rho_t)\,dW_t,
\end{multline*}
where $[A,B]=AB-BA$, has global Lipschitz coefficients and hence has a 
unique strong solution in $\mathbb{C}^{N\times N}$ and a.s.\ continuous
adapted sample paths \cite{protter}.  Moreover $\bar\rho_t$ must be 
bounded as $C(\rho)$ has compact support.  Hence $U_t=U(\bar\rho_t)$ is an 
a.s.\ continuous, bounded adapted process.

Now consider the solution $\rho_t$ of (\ref{eq:qfilt}) with
$u_t=U(\bar\rho_t)$ and $\rho_0=\bar\rho_0\in\mathcal{S}$.  As both
$\rho_t$ and $\bar\rho_t$ have a unique solution, the solutions must
coincide up to the first exit time from $\mathcal{T}$.  But we have
already established that $\rho_t$ remains in $\mathcal{S}$ for all $t>0$,
so $\bar\rho_t$ can certainly never exit $\mathcal{T}$.  Hence
$\bar\rho_t=\rho_t$ for all $t>0$, and the result follows. \qquad
\end{proof}

In the following, we will denote by $\varphi_t(\rho,u)$ the solution of
(\ref{eq:qfilt}) at time $t$ with the control $u_t=u(\rho_t)$ and initial
condition $\rho_0=\rho\in\mathcal{S}$.

\begin{proposition}
\label{pro:feller}
    If $V(\rho)$ is continuous, then $\mathbb{E}V(\varphi_t(\rho,u))$
    is continuous in $\rho$; i.e., the diffusion {\rm (\ref{eq:qfilt})} is
    Feller continuous.
\end{proposition}

\begin{proof}
    Let $\{\rho^n\in\mathcal{S}\}$ be a sequence of points converging to
    $\rho^\infty\in\mathcal{S}$.  Let us write
    $\rho^n_t=\varphi_t(\rho^n,u)$ and
    $\rho^\infty_t=\varphi_t(\rho^\infty,u)$.
    First, we will show that
    \begin{equation}
    \label{eq:toprove1feller}
        \EE\|\rho^n_t-\rho^\infty_t\|^2 \rightarrow 0
        \quad \mbox{as} \quad n \rightarrow \infty.
    \end{equation}
    where $\|\cdot\|$ is the Frobenius norm ($\|A\|^2=(A,A)$ with
    the inner product $(A,B)=\tr{A^*B}$).  We will write
    $\delta_t^n=\rho^n_t-\rho^\infty_t$.  Using It\^o's rule we obtain
    \begin{equation}
    \label{eq:estimatesfeller}
    \begin{split}
        \EE\|\delta_t^n&\|^2 =
        \|\delta_0^n\|^2
        +\int_0^t \eta\EE\tr{(c\delta_s^n+\delta_s^nc^*
            -{\rm Tr}[(c+c^*)\rho_s^n]\rho_s^n
            +{\rm Tr}[(c+c^*)\rho_s^\infty]\rho_s^\infty)^2}ds
             \\&
        +\int_0^t 2\,\EE\left[
        \tr{(i[\rho_s^n,H(\rho_s^n)]
            -i[\rho_s^\infty,H(\rho_s^\infty)])\delta_s^n}
        +\tr{c\delta_s^n c^*\delta_s^n-c^*c(\delta_s^n)^2}
        \right]ds
    \end{split}
    \end{equation}
    where $[A,B]=AB-BA$.  Let us estimate each of these terms.  We have
    \begin{equation}
    \begin{split}
        \tr{c^*c(\delta_t^n)^2} &=
            \|c\delta_t^n\|^2\le C_1\|\delta_t^n\|^2 \\
        \tr{c\delta_t^n c^*\delta_t^n} &=
            (\delta_t^n c,c\delta_t^n)\le
            \|\delta_t^n c\|~\|c\delta_t^n\|\le C_2\|\delta_t^n\|^2
    \end{split}
    \end{equation}
    where we have used the Cauchy-Schwartz inequality and the fact
    that all the operators are bounded.  Next we tackle
    \begin{equation}
        \tr{(i[\rho_t^n,H(\rho_t^n)]
            -i[\rho_t^\infty,H(\rho_t^\infty)])\delta_t^n}\le
        \|i[\rho_t^n,H(\rho_t^n)]-i[\rho_t^\infty,H(\rho_t^\infty)]\|
            ~\|\delta_t^n\|.
    \end{equation}
    Now note that $S(\rho)=i[\rho,H(\rho)]=i[\rho,F+u(\rho)G]$ is
    $C^1$ in the matrix elements of $\rho$, and its derivatives
    are bounded as $\mathcal{S}$ is compact.  Hence $S(\rho)$ is
    Lipschitz continuous, and we have
    \begin{equation}
        \|S(\rho_t^n)-S(\rho_t^\infty)\|\le C_3\|\rho_t^n-
            \rho_t^\infty\|=C_3\|\delta_t^n\|
    \end{equation}
    which implies
    \begin{equation}
        \tr{(i[\rho_t^n,H(\rho_t^n)]
            -i[\rho_t^\infty,H(\rho_t^\infty)])\delta_t^n}\le
            C_3\|\delta_t^n\|^2.
    \end{equation}
    Finally, we have $\|c\delta_t^n+\delta_t^nc^*\|\le C_4\|\delta_t^n\|$
    due to boundedness of multiplication with $c$, and a similar
    Lipschitz argument as the one above can be applied to
    $S'(\rho)={\rm Tr}[(c+c^*)\rho]\rho$, giving
    \begin{equation}
        \|{\rm Tr}[(c+c^*)\rho_t^n]\rho_t^n
        -{\rm Tr}[(c+c^*)\rho_t^\infty]\rho_t^\infty\|
        \le C_5\|\delta_t^n\|.
    \end{equation}
    We can now use $\|A+B\|^2\le\|A\|^2+2\|A\|\,\|B\|+\|B\|^2$ to estimate
    the last term in (\ref{eq:estimatesfeller}) by $C_6\|\delta_t^n\|^2$.
    Putting all these together, we obtain
    \begin{equation}
        \EE\|\delta_t^n\|^2 \le
            \|\delta_0^n\|^2+C\int_0^t \EE\|\delta_s^n\|^2 ds
    \end{equation}
    and thus by Gronwall's lemma
    \begin{equation}
    \EE\|\delta_t^n\|^2 \leq e^{Ct}\|\delta_0^n\|^2=
        e^{Ct}\|\rho^n-\rho^\infty\|^2.
    \end{equation}
    As $t$ is fixed, Eq.\ (\ref{eq:toprove1feller}) follows.

    We have now proved that $\rho_t^n\to\rho_t^\infty$ in mean square
    as $n\to\infty$, which implies convergence in probability.
    But then for any continuous $V$, $V(\rho_t^n)\to V(\rho_t^\infty)$
    in probability (\cite{gikhman}, pp.\ 60).  As $\mathcal{S}$ is compact,
    $V$ is bounded and we have
    \begin{equation}
        \mathbb{E}V(\rho_t^\infty)=
        \mathbb{E}[
         \mathop{\mathbb{P}\mbox{-lim}}_{n\to\infty}
        V(\rho_t^n)]=
        \lim_{n\to\infty}\mathbb{E}V(\rho_t^n)
    \end{equation}
    by dominated convergence (\cite{gikhman}, pp.\ 72).
    But as this holds for any convergent sequence $\rho^n$, the
    result follows.
    \qquad
\end{proof}

\begin{proposition}
\label{pro:markov}
    $\varphi_t(\rho,u)$ is a strong Markov process in $\mathcal{S}$.
\end{proposition}

\begin{proof}
    The proof of the Markov property in \cite{oksendal},
    pp.\ 109--110, carries over to our case.  But then
    the strong Markov property follows from
    Feller continuity \cite{kushner-67}.
    \qquad
\end{proof}

\begin{proposition}
\label{pro:stopopen}
    Let $\tau$ be the first exit time of $\rho_t$ from an
    open set $Q\subset\mathcal{S}$ and consider the stopped
    process $\rho_t^Q=\varphi_{t\wedge\tau}(\rho,u)$.  Then $\rho_t^Q$
    is also a strong Markov process in $\mathcal{S}$.
    Furthermore, for $V$ s.t.\ $\mathscr{A}V$ exists and is continuous,
    where $\mathscr{A}$ is the weak infinitesimal operator associated to
    $\varphi_t(\rho,u)$, we have $\mathscr{A}_QV(x)=\mathscr{A}V(x)$ if
    $x\in Q$ and $\mathscr{A}_QV(x)=0$ if $x\ne Q$ for the weak
    infinitesimal operator $\mathscr{A}_Q$ associated to $\rho_t^Q$.
\end{proposition}

\begin{proof}
    This follows from \cite{kushner-67}, pp.\ 11--12, and Proposition
    \ref{pro:uniformstoch}. \qquad
\end{proof}

\section{Angular momentum systems}
\label{single:sec}

In this section we consider a quantum system with fixed angular momentum $J$
($2J\in\mathbb{N}$), e.g.\ an atomic ensemble, which is detected through a
dispersive optical probe \cite{vanhandel-review}.  After conditioning,
such systems are described by an equation of the form (\ref{eq:qfilt}) where
\begin{itemize}
\item The Hilbert space dimension $N=2J+1$;
\item $c=\beta F_z$, $F=0$ and $G=\gamma F_y$ with $\beta,\gamma>0$.
\end{itemize}
Here $F_y$ and $F_z$ are the (self-adjoint) angular momentum operators
defined as follows.   Let $\{\psi_k:k=0\ldots 2J\}$ be the standard basis in
$\mathbb{C}^N$, i.e.\ $\psi_i$ is the vector with a single nonzero
element $\psi_i^i=1$.  Then \cite{merzbacher}
\begin{equation}
\begin{split}
    F_y\psi_k &= ic_{k-J}\psi_{k+1}-ic_{J-k}\psi_{k-1}, \\
    F_z\psi_k &= (k-J)\psi_k
\end{split}
\end{equation}
with $c_m=\tfrac{1}{2}\sqrt{(J-m)(J+m+1)}$.  Without loss of generality
we will choose $\beta=\gamma=1$, as we can always rescale time and $u_t$
to obtain any $\beta,\gamma$.

Let us begin by studying the dynamical behavior of the resulting
equation,
\begin{equation}
\label{single:eq}
    d\rho_t=-iu_t[F_y,\rho_t]\,dt
     -\tfrac{1}{2}[F_z,[F_z,\rho_t]]\,dt
     +\sqrt{\eta}\,(F_z\rho_t+\rho_t F_z
        -2\,{\rm Tr}[F_z\rho_t]\rho_t)\,dW_t
\end{equation}
without feedback $u_t=0$.

\begin{proposition}[Quantum state reduction]
\label{pro:reduction}
    For any $\rho_0\in\mathcal{S}$, the solution $\rho_t$
    of {\rm (\ref{single:eq})} with $u_t=0$ converges a.s.\ as $t\to\infty$
    to one of $\psi_m\psi_m^*$.
\end{proposition}

\begin{proof}
    We will apply Theorem \ref{thm:localstab} with $Q_\lambda=\mathcal{S}$.  
    Consider the Lyapunov function $v(\rho)=\mbox{Tr}[F_z^2\rho]-
    (\mbox{Tr}[F_z\rho])^2$.  One easily calculates $\mathscr{A}v(\rho)=
    -4\eta\,v(\rho)^2\le 0$ and hence
    \begin{equation}
        \mathbb{E}v(\rho_t)=v(\rho_0)
            -4\eta\int_0^t\mathbb{E}v(\rho_s)^2\,ds
    \end{equation}
    by using the It\^o rules.  Note that $v(\rho)\ge 0$, so
    \begin{equation}
        4\eta\int_0^t\mathbb{E}v(\rho_s)^2\,ds
        =v(\rho_0)-\mathbb{E}v(\rho_t)\le v(\rho_0)<\infty.
    \end{equation}
    Thus we have by monotone convergence
    \begin{equation}\label{eq:mononoco}
        \mathbb{E}\int_0^\infty v(\rho_s)^2\,ds<\infty \quad
	\Longrightarrow \quad \int_0^\infty v(\rho_s)^2\,ds<\infty
	\quad\mbox{a.s.}
    \end{equation}
    By Theorem \ref{thm:localstab} the limit of $v(\rho_t)$ as 
    $t\to\infty$ exists a.s., and hence Eq.\ (\ref{eq:mononoco}) implies 
    that $v(\rho_t)\to 0$ a.s.  But the only states $\rho$ that satisfy
    $v(\rho)=0$ are $\rho=\psi_m\psi_m^*$.
    \qquad
\end{proof}

The main goal of this section is to provide a feedback control
law that globally stabilizes \eqref{single:eq} around the equilibrium
solution $(\rho_t\equiv\rho_f,u\equiv0)$, where we select a target state
$\rho_f=v_fv_f^*$ from one of $v_f=\psi_m$.

Stabilization of quantum state reduction for low-dimensional
angular momentum systems has been studied in~\cite{vanhandel-05}.
It is shown that the main challenge in such a stabilization
problem is due to the geometric symmetry hidden in the state space
of the system. Many natural feedback laws fail to stabilize the
closed-loop system around the equilibrium point $\rho_f$ because
of this symmetry: the $\omega$-limit set contains points other
than $\rho_f$. The approach of~\cite{vanhandel-05} uses computer
searches to find continuous control laws that break this symmetry
and globally stabilize the desired state.  Unfortunately, the
method is computationally involved and can only be applied to
low-dimensional systems.  Additionally, it is difficult to prove
stability in this way for arbitrary parameter values, as the
method is not analytical.

Here we present a different approach which avoids the unwanted
limit points by changing the feedback law around them.  The approach is
entirely analytical and globally stabilizes the desired target state for
any dimension $N$ and $0<\eta\le 1$.  The main result of this section can
be stated as follows:

\begin{theorem}
\label{single:thm}
Consider the system~\eqref{single:eq} evolving in the set $\SSS$.
Let $\rho_f=v_fv_f^*$ where $v_f$ is one of $\psi_m$, and let
$\gamma>0$.  Consider the following control law:
\begin{enumerate}
    \item $u_t=-\tr{i[F_y,\rho_t]\rho_f}$ if $\tr{\rho_t \rho_f} \geq \gamma$;
    \item $u_t=1$ if $\tr{\rho_t \rho_f} \leq \gamma/2 $;
    \item If $\rho_t\in\mathcal{B}=\{\rho:\gamma/2<\tr{\rho\rho_f}<\gamma\}$,
    then $u_t=-\tr{i[F_y,\rho_t]\rho_f}$ if $\rho_t$
    last entered $\mathcal{B}$ through the boundary
    $\tr{\rho\rho_f}=\gamma$, and $u_t=1$ otherwise.
\end{enumerate}
Then $\exists\gamma>0$ s.t.\ $u_t$ globally stabilizes
\eqref{single:eq} around $\rho_f$ and $\mathbb{E}\rho_t\to\rho_f$ as
$t\to\infty$.
\end{theorem}

Throughout the proofs we use the ``natural'' distance function
$$
    V(\rho)=1-\tr{\rho\rho_f}:\SSS\rightarrow[0,1]
$$
from the state $\rho$ to the target state $\rho_f$.  For future reference,
let us define for each $\alpha\in[0,1]$ the level set $\SSS_\alpha$ to be
$$
    \SSS_\alpha=\{\rho\in\SSS:V(\rho)=\alpha\}.
$$
Furthermore, we define the following sets:
\begin{equation*}
\begin{split}
    \SSS_{>\alpha} &= \{\rho\in\SSS : \alpha<V(\rho)\leq 1\}, \\
    \SSS_{\ge\alpha} &= \{\rho\in\SSS : \alpha\le V(\rho)\leq 1\}, \\
    \SSS_{<\alpha} &= \{\rho\in\SSS : 0\le V(\rho)< \alpha\}, \\
    \SSS_{\le\alpha} &= \{\rho\in\SSS : 0\le V(\rho)\leq \alpha\}. \\
\end{split}
\end{equation*}
The proof of Theorem~\ref{single:thm} proceeds in four steps:
\begin{enumerate}
\item In the first step we show that when the initial state lies in the
set $\SSS_1$, the constant control field $u=1$ ensures the exit of the
trajectories (at least) in expectation from the level set $\SSS_1$.
\item In the second step we use the result of step 1 to show that there
exists a $\gamma>0$ such that whenever the initial state lies inside the
set $\SSS_{>1-\gamma}$ and the control field is taken to be $u=1$, the 
expectation value of the first exit time from this set takes a finite 
value.  Thus if we start the controlled system in the set 
$\SSS_{>1-\gamma}$, it will exit this set in finite time with probability 
one.
\item In the third step we show that whenever the initial state lies
inside the set $\SSS_{\le 1-\gamma}$ and the control is given by the
feedback law $u(t)=-\tr{i[F_y,\rho_t]\rho_f}$, the sample paths never exit 
the set $\SSS_{<1-\gamma/2}$ with a probability uniformly larger than a 
strictly positive value.  We also show that almost all paths that never 
leave $\SSS_{<1-\gamma/2}$ converge to the equilibrium point $\rho_f$.
\item In the final step, we prove that there is a unique solution $\rho_t$ 
under the control $u_t$ by piecing together the solutions with fixed 
controls $u=1$ and $u=-\tr{i[F_y,\rho_t]\rho_f}$.  Combining the results 
of the second and the third step, we show that the resulting trajectories 
of the system eventually converge toward the equilibrium state $\rho_f$ 
with probability one.
\end{enumerate}

\subsection*{Step 1}

Let us take a fixed time $T>0$ and define the nonnegative function
$$
    \chi(\rho)=\min_{t\in[0,T]}\EE V(\varphi_t(\rho,1)),\qquad
    \rho \in \SSS.
$$
Recall that $\varphi_t(\rho,1)$ denotes the solution of (\ref{single:eq})
at time $t$ with the control $u_t=1$ and initial condition $\rho_0=\rho$.
The goal of the first step is to show the following result:

\begin{lemma}\label{first:lem}
$\chi(\rho)<1 ~~ \forall \rho\in \SSS_1.$
\end{lemma}

To prove this statement we will first show the following deterministic
result.

\begin{lemma}\label{add:lem}
Consider the deterministic differential equation
\begin{equation}\label{det:eq}
    \frac{d}{dt}v_t=(-i F_y-F_z^2+CF_z)v_t,\qquad
        v_0\in\CC^N\setminus\{0\}.
\end{equation}
For sufficiently large $C\gg 1$, $v_t$ exits the set
$\{v:v^*v_f=0\}$ in the interval $[0,T]$, i.e.\ there exists
$t\in[0,T]$ such that $v_t^*v_f\neq 0$.
\end{lemma}

\begin{proof}
The matrices $F_z$ and $F_y$ are of the form
$$
 F_z =
  \begin{pmatrix}
    * & &&  & 0 \\
    & * &&& \\
     & &\ddots&& \\
    &&&*& \\
     0 & && & * \\
  \end{pmatrix}, \qquad
 F_y =
  \begin{pmatrix}
    0 & * & & &0 \\
    * & 0 &*&  &\\
      & \ddots &\ddots &\ddots&   \\
      &        &*  &0  & *\\
    0 &        &  & *&0\\
  \end{pmatrix}
$$
where $F_z$ has no repeated diagonal entries ($F_z$ has a nondegenerate
spectrum) and the starred elements directly above and below the diagonal
of $F_y$ are all nonzero.

Now choose a constant $\kappa$ so that the matrix
$$
    A=-iF_y-F_z^2+\kappa F_z
$$
admits distinct eigenvalues.  This is always possible by choosing
sufficiently large $\kappa$, as $F_z$ has nondegenerate eigenvalues and
the eigenvalues of $A$ depend continuously\footnote{
    Note that the coefficients of the characteristic polynomial of $A$
    are continuous functions of $\kappa$, and the roots of a
    polynomial depend continuously on the polynomial coefficients.
} on $\kappa$.  For $k\in\{1,..,N\}$ define the matrices $A_{k-1}$ and
$\tilde A_{k+1}$ to be:
$$
A_{k-1}=[A_{ij}]_{1\leq i,j\leq k-1},\qquad \tilde
A_{k+1}=[A_{ij}]_{k+1\leq i,j\leq N}.
$$
The fact that the matrices $[(F_z)_{ij}]_{1\leq i,j \leq k-1}$ and
$[(F_z)_{ij}]_{k+1\leq i,j \leq N}$ have different eigenvalues then imply
that for sufficiently large $\kappa$ the matrices $A_{k-1}$ and $\tilde
A_{k+1}$ have disjoint spectra as well.

Suppose that the solution of
$$
    \dot v= A v, \qquad v|_{t=0}=v_0
$$
never leaves the set $\{v:v^*v_f=0\}$ in the interval $t\in[0,T]$.
Then in particular
$$
    \frac{d^n}{dt^n}v^*v_f|_{t=0}=(A^nv_0)^*v_f=0,\qquad
    n=0,1,\ldots
$$
The matrix $A$ is diagonalizable as it has distinct eigenvalues, i.e.\
$A=P D P^{-1}$ where $D$ is a diagonal matrix. Thus
\begin{equation}\label{deriv:eq}
    (D^n \tilde v_0)^*\tilde v_f=0, \qquad n=0,1,\ldots
\end{equation}
where $\tilde v_0=P^{-1}v_0$ and $\tilde v_f=P^*v_f$. Eq.\
\eqref{deriv:eq} implies that $M\tilde v_0=0$ where
$$
 M =
  \begin{pmatrix}
    (\tilde v_f)_1^* &   & \ldots & (\tilde v_f)_N^* \\
    (\tilde v_f)_1^* D_{11}  &   &\ldots&  (\tilde v_f)_N^* D_{NN}\\
    (\tilde v_f)_1^* D_{11}^{2}&  &\ldots& (\tilde v_f)_N^* D_{NN}^{2} \\
     \vdots &   &   \vdots   &  \vdots   \\
     (\tilde v_f)_1^* D_{11}^{N-1}  &
      &\ldots& (\tilde v_f)_N^* D_{NN}^{N-1} \\
  \end{pmatrix}.
$$
The determinant of this Vandermonde matrix is
$$
    {\rm det}\,M=
        (\tilde v_f)_1^*\cdots(\tilde v_f)_N^*
        \prod_{i>j}(D_{ii}-D_{jj}).
$$
As the matrix $A$ has distinct eigenvalues, all the entries
$D_{11},D_{22},...,D_{NN}$ are different. Thus if we can show that
all the entries of the vector $\tilde v_f$ are non-zero then the
matrix $M$ must be invertible.  But then $M\tilde v_0=0$ implies
that $\tilde v_0=0$ and hence $v_0=0$ is the only initial state
for which the dynamics does not leave the set $\{v:v^*v_f=0\}$ in
the interval $t\in[0,T]$, proving our assertion.

Let us thus show that in fact all elements of $\tilde v_f$ are nonzero.
Note that
$$
(\tilde v_f)_k=(P^*v_f)_k=P_{fk}^*,
$$
so it suffices to show that the eigenvectors of the matrix $A$ have
only nonzero elements. Suppose that an eigenvector $\Xi$ of $A$ admits
a zero entry, i.e.
$$
    A\Xi=\lambda\Xi,\qquad \Xi_k=0 \text{  for some  }
    k\in\{1,..,N\}.
$$
Defining $\chi_{k-1}=[\Xi_j]_{j=1,..,k-1}$ and $\tilde
\chi_{k+1}=[\Xi_j]_{j=k+1,..,N}$, a straightforward computation
shows that due to the structure of the matrix $A$
$$
    A_{k-1}\chi_{k-1}=\lambda\chi_{k-1}\quad \text{and} \quad \tilde
    A_{k+1}\tilde\chi_{k+1}=\lambda\tilde \chi_{k+1}.
$$
But by the discussion above $A_{k-1}$ and $\tilde A_{k+1}$ have
disjoint spectra, so $\Xi$ can only be an eigenvector if either
$\chi_{k-1}=0$ or $\tilde\chi_{k+1}=0$.

Let us consider the case where $\chi_{k-1}=0$; the treatment of the second
case follows an identical argument.  Let $j > k$ be the first non-zero
entry of $\Xi$, i.e.\
\begin{equation}\label{zero:eq}
    \Xi_1=\Xi_2=...=\Xi_{j-1}=0\quad \text{and} \quad \Xi_j\neq 0.
\end{equation}
As $A\Xi=\lambda\Xi$, we have that
$$
    0=\lambda\Xi_{j-1}=
    A_{j-1,j-2}\Xi_{j-2}+A_{j-1,j-1}\Xi_{j-1}
    +A_{j-1,j}\Xi_j=A_{j-1,j}\Xi_j
    =-i(F_y)_{j-1,j}\Xi_j.
$$
As $(F_y)_{j-1,j}\neq 0$ this relation ensures that $\Xi_j=0$. But this is
in contradiction with~\eqref{zero:eq} and so $\Xi$ cannot admit any zero
entry.  This completes the proof. \qquad
\end{proof}

{\em Proof of Lemma \ref{first:lem}}.
We begin by restating the problem as in the proof of Lemma
\ref{lem:invzakai}.  We can write $\varphi_t(\rho,1)=
\tilde\rho_t/{\rm Tr}\,\tilde\rho_t$ with $\tilde\rho_t=
\sum_i\lambda_i\mathbb{E}[v_t^iv_t^{i*}|\mathcal{F}_t^y]$, where
$\lambda_i$ are convex weights and $v_t^i$ are given by the equations
\begin{equation}\label{eq:zaksupp}
    dv_t^i=-iF_yv_t^i\,dt-\tfrac{1}{2}F_z^2v_t^i\,dt+
        F_zv_t^i\,dW_t',\qquad v_0^i\in\CC^N\setminus\{0\}.
\end{equation}
Note that $\mathbb{E}{\rm Tr}[\varphi_t(\rho,1)\rho_f]=0$ iff
$\mathbb{E}{\rm Tr}[\tilde\rho_t\rho_f]=
\sum_i\lambda_i\mathbb{E}[v_t^{i*}\rho_fv_t^i]=0$.  But as
$v_t^{i*}\rho_fv_t^i\ge 0$, we obtain $\EE V(\varphi_t(\rho,1))=1$
iff $v_t^{i*}v_f=0$ a.s.\ for all $i$.

To prove the assertion of the Lemma, it suffices to show that there exists
a $t\in[0,T]$ such that $\EE V(\varphi_t(\rho,1))<1$.  Thus it is
sufficient to prove that
\begin{equation}\label{exitp2:eq}
    \exists t\in[0,T]\quad\mbox{s.t.}\quad\PP(v_t^{*}v_f\ne 0)>0
\end{equation}
where $v_t$ is the solution of an equation of the form (\ref{eq:zaksupp}).
To this end we will use the support theorem, Theorem \ref{thm:supportth},
together with Lemma \ref{add:lem}.

To apply the support theorem we must first take care of two preliminary
issues.  First, the support theorem in the form of Theorem
\ref{thm:supportth} must be applied to stochastic differential equations
with a Wiener process as the driving noise, whereas the noise $W_t'$ of
Eq.\ (\ref{eq:zaksupp}) is a Wiener process with (bounded) drift:
\begin{equation}
	dW_t'=\sqrt{\eta}\,dy_t+\sqrt{1-\eta}\,d\hat W_t=
		2\eta\,{\rm Tr}[F_z\rho_t]dt
		+\sqrt{\eta}\,dW_t+\sqrt{1-\eta}\,d\hat W_t.
\end{equation}
Using Girsanov's theorem, however, we can find a new measure $\mathbb{Q}$ 
that is equivalent to $\mathbb{P}$, such that $W_t'$ is a Wiener process 
under $\mathbb{Q}$ on the interval $[0,T]$.  But as the two measures are
equivalent,
\begin{equation}\label{exitp:eq}
    \exists t\in[0,T]\quad\mbox{s.t.}\quad\mathbb{Q}(v_t^{*}v_f\ne 0)>0
\end{equation}
implies (\ref{exitp2:eq}).  Second, the support theorem refers to an
equation in the Stratonovich form; however, we can easily find the
Stratonovich form
\begin{equation}\label{eq:suppstrat}
    dv_t=-iF_yv_t\,dt-F_z^2v_t\,dt+
        F_zv_t\circ dW_t'
\end{equation}
which is equivalent to (\ref{eq:zaksupp}).  It is easily verified that
this linear equation satisfies all the requirements of the support
theorem.

To proceed, let us suppose that~\eqref{exitp:eq} does not hold true.
Then
\begin{equation}\label{cont:eq}
    \mathbb{Q}(v_t^{*}v_f=0)=1\qquad
    \forall t\in[0,T].
\end{equation}
Recall the following sets: $\mathscr{W}_{v_0}$ is the set of continuous
paths starting at $v_0$, and $\mathscr{S}_{v_0}$ is the smallest closed
subset of $\mathscr{W}_{v_0}$ such that
$\mathbb{Q}(\{\omega\in\Omega:v_\cdot(\omega)\in\mathscr{S}_{v_0}\})=1$.
Now denote by $\mathscr{T}_{v_0,t}$ the subset of $\mathscr{W}_{v_0}$ such
that $v_t^{*}v_f=0$, and note that $\mathscr{T}_{v_0,t}$ is closed in the
compact uniform topology for any $t$.  Then (\ref{cont:eq}) would imply
that $\mathscr{S}_{v_0}\subset\mathscr{T}_{v_0,t}$ for all $t\in[0,T]$.
But by the support theorem the solutions of (\ref{det:eq}) are elements of
$\mathscr{S}_{v_0}$, and by Lemma \ref{add:lem} there exists a time
$t\in[0,T]$ and a constant $C$ such that the solution of (\ref{det:eq}) is
not an element of $\mathscr{T}_{v_0,t}$.  Hence we have a contradiction,
and the assertion is proved.
\qquad\endproof

\subsection*{Step 2}

We begin by extending the result of Lemma \ref{first:lem} to hold
uniformly in a neighborhood of the level set $\mathcal{S}_1$.

\begin{lemma}\label{fourth:lem}
There exists $\gamma>0$ such that $\chi(\rho)<1-\gamma$ for all
$\rho\in\SSS_{\ge 1-\gamma}$.
\end{lemma}

\begin{proof}
Suppose that for every $\xi>0$ there exists a matrix
$\rho_\xi\in\SSS_{>1-\xi}$ such that
$$
    1-\xi<\chi(\rho_\xi)\leq 1.
$$
By extracting a subsequence $\xi_n \searrow 0$ and using the
compactness of $\SSS$, we can assume that $\rho_{\xi_n}\rightarrow
\rho_\infty\in \SSS_1$ and that $\chi(\rho_{\xi_n})\rightarrow 1$.
But by Lemma~\ref{first:lem} $\chi(\rho_\infty)=1-\epsilon<1$.
Now choose $s\in[0,T]$ such that
$$
    \EE V(\varphi_s(\rho_\infty,1))=1-\epsilon.
$$
Using Feller continuity, Prop.\ \ref{pro:feller}, we can now write
$$
    1=\lim_{n\rightarrow\infty}\chi(\rho_{\xi_n})
    \leq\lim_{n\rightarrow \infty}\EE
        V(\varphi_s(\rho_{\xi_n},1))
    =\EE V(\varphi_s(\rho_\infty,1))
    =1-\epsilon<1.
$$
which is a contradiction.  Hence there exists $\xi>0$ such that
$\chi(\rho)\le 1-\xi$ for all $\rho\in\mathcal{S}_{>1-\xi}$.
The result follows by choosing $\gamma=\xi/2$.
\qquad
\end{proof}

The following Lemma is the main result of the second step.

\begin{lemma}\label{fifth:lem}
Let $\tau_{\rho}(\mathcal{S}_{>1-\gamma})$ be the first exit time of
$\varphi_t(\rho,1)$ from $\mathcal{S}_{>1-\gamma}$.  Then
$$
    \sup_{\rho\in\mathcal{S}_{>1-\gamma}}
    \EE\tau_{\rho}(\mathcal{S}_{>1-\gamma})<\infty.
$$
\end{lemma}

\begin{proof}
The following result can be found in Dynkin (\cite{dynkin-book1}, pp.\
111, Lemma 4.3):
$$
    \EE\tau_\rho(\SSS_{>1-\gamma})\leq
    \frac{T}{
    1-\sup_{\zeta\in\SSS}\PP\{\tau_\zeta(\SSS_{>1-\gamma})> T\}}.
$$
We will show that
\begin{equation}\label{prob:eq}
    \sup_{\zeta \in \SSS}
    \PP\{\tau_\zeta(\SSS_{>1-\gamma})>T\}<1.
\end{equation}
This holds trivially for $\zeta\in\SSS_{\le 1-\gamma}$, as then
$\tau_\zeta(\SSS_{>1-\gamma})=0$.  Let us thus suppose that
$$
    \forall \epsilon>0 \quad
    \exists \zeta_{\epsilon}\in\SSS_{>1-\gamma}
        \quad \text{such that} \quad
    \PP\{\tau_{\zeta_\epsilon}(\SSS_{>1-\gamma})>T\}>1-\epsilon.
$$
Then for all $s\in[0,T]$, we have that
$$
    \EE V(\varphi_s(\zeta_\epsilon,1))
    > (1-\epsilon)\inf_{\rho\in\SSS_{>1-\gamma}}V(\rho)
    =(1-\epsilon)(1-\gamma).
$$
By compactness there exists a sequence $\epsilon_n \searrow 0$ and
$\zeta_\infty \in \SSS_{\ge 1-\gamma}$ such that
$\zeta_{\epsilon_n}\rightarrow \zeta_\infty$ as $n\rightarrow
\infty$. Thus by Prop.\ \ref{pro:feller}
$$
    \EE V(\varphi_s(\zeta_\infty,1))> 1-\gamma \quad \forall s\in[0,T].
$$
But this is in contradiction with result of Lemma~\ref{fourth:lem}.
Hence there exists an $\epsilon>0$ such that
$\sup_{\zeta\in\SSS}\PP\{\tau_{\zeta}(\SSS_{>1-\gamma})>T\}=
1-\epsilon$, and we obtain
$$
    \EE(\tau_{\rho}(\SSS_{>1-\gamma}))\leq
    \frac{T}{1-(1-\epsilon)}
    =\frac{T}{\epsilon}<\infty
$$
uniformly in $\rho$.  This completes the proof.  \qquad
\end{proof}

\subsection*{Step 3}

In this step we deal with the situation where the initial state lies
inside the set $\SSS_{\le 1-\gamma}$.  We will denote by
$u_1(\rho)=-\tr{i[F_y,\rho]\rho_f}$ and by $\varphi_t(\rho,u_1)$ the
solution of (\ref{single:eq}) with $\rho_0=\rho$ and with
$u_t=u_1(\rho_t)$.  Denote by $\mathscr{A}$ the weak infinitesimal
operator of $\varphi_{t}(\rho,u_1)$.  We will apply the stochastic
Lyapunov theorems with $Q_\lambda=\mathcal{S}$.

We begin by showing that there is a non-zero probability $p>0$ that
whenever the initial state lies inside $\SSS_{\le 1-\gamma}$ the
trajectories of the system never exit the set $\SSS_{<1-\gamma/2}$.

\begin{lemma}\label{sixth:lem}
For all $\rho \in \SSS_{\le 1-\gamma}$
$$
    \PP\left[
        \sup_{0\le t<\infty}V(\varphi_t(\rho,u_1))
        \geq 1-\gamma/2
    \right]
    \leq 1-p=\frac{1-\gamma}{1-\gamma/2}<1.
$$
\end{lemma}

\begin{proof}
This follows from Theorem \ref{thm:localstab} and
$\mathscr{A}V(\rho)=-u_1(\rho)^2\leq 0$.
\qquad
\end{proof}

We now restrict ourselves to the paths that never leave
$\SSS_{<1-\gamma/2}$. We will first show that these paths converge toward
$\rho_f$ in probability.  We then extend this result to prove almost
sure convergence.

\begin{lemma}\label{seventh2:lem}
    The sample paths of $\varphi_t(\rho,u_1)$ that never exit the set
    $\SSS_{<1-\gamma/2}$ converge in probability to $\rho_f$ as
    $t\to\infty$.
\end{lemma}

\begin{proof}
Consider the Lyapunov function
$$
    \VV(\rho)=1-\tr{\rho\rho_f}^2.
$$
It is easily verified that $\VV(\rho)\geq 0$ for all $\rho\in\SSS$ and
that $\VV(\rho)=0$ iff $\rho=\rho_f$. A straightforward computation gives
\begin{equation*}
    \mathscr{A}\VV(\rho)=
    -2u_1(\rho)^2\,\tr{\rho\rho_f}
    -4\eta\,(\lambda_f-\tr{\rho F_z})^2\,\tr{\rho\rho_f}^2
    \le 0
\end{equation*}
where $\lambda_f$ is the eigenvalue of $F_z$ associated to $v_f$.  Now
note that all the conditions of Theorem \ref{thm:lasalle} are satisfied by
virtue of Prop.\ \ref{pro:feller} and \ref{pro:uniformstoch}.  Hence
$\varphi_{t}(\rho,u_1)$ converges in probability to the largest invariant
set contained in $\mathcal{C}=\{\rho\in\mathcal{S}:\mathscr{A}\VV(\rho)=0\}$.

In order to satisfy the condition $\mathscr{A}\VV(\rho)=0$, we must have
$u_1(\rho)^2\,\tr{\rho\rho_f}=0$ as well as $(\lambda_f-\tr{\rho
F_z})^2\,\tr{\rho\rho_f}^2=0$.
The latter implies that
$$
    \text{either}\quad \tr{\rho\rho_f}=0 \qquad \text{or} \quad
    \tr{\rho F_z}=\lambda_f.
$$
Let us investigate the largest invariant set contained in
$\mathcal{C}'=\{\rho\in\SSS:\tr{\rho F_z}=\lambda_f\}$.  Clearly this
invariant set can only contain $\rho\in\mathcal{C}'$ for which
$\tr{\varphi_{t}(\rho,u_1)F_z}$ is constant.  Using It\^o's rule we obtain
$$
    d\,\tr{\rho_tF_z}=-iu_1(\rho_t)\,\tr{[F_y,\rho_t]F_z}\,dt
        +2\sqrt{\eta}\,(\tr{F_z^2\rho_t}-\tr{F_z\rho_t}^2)\,dW_t.
$$
Hence in order for $\tr{\varphi_{t}(\rho,u_1)F_z}$ to be constant, we must
at least have
$$
    \tr{F_z^2\rho}-\tr{F_z\rho}^2=0.
$$
But as in the proof of Prop.\ \ref{pro:reduction}, this implies that
$\rho=\psi_m\psi_m^*$ for some $m$, and thus the only possibilities are
$V(\rho)=0$ (for $\rho=v_fv_f^*$) or $V(\rho)=1$.

From the discussion above it is evident that the largest
invariant set contained in $\mathcal{C}$ must be contained inside the set
$\{\rho_f\}\cup\mathcal{S}_1$.  But then the paths that never exit
$\mathcal{S}_{<1-\gamma/2}$ must converge in probability to $\rho_f$.
Thus the assertion is proved.
\qquad
\end{proof}

\begin{lemma}\label{seventh:lem}
    $\varphi_t(\rho,u_1)$ converges to $\rho_f$ as $t\to\infty$ for
    almost all paths that never exit the set $\SSS_{<1-\gamma/2}$.
\end{lemma}

\begin{proof}
Define the event $P^\rho_{<1-\gamma/2}=\{\omega\in\Omega:
\varphi_t(\rho,u_1)\mbox{ never exits }\SSS_{<1-\gamma/2}\}$.
Then Lemma \ref{seventh2:lem} implies that
$$
    \lim_{t\to\infty}
    \mathbb{P}\left(\|\varphi_t(\rho,u_1)-\rho_f\|>\varepsilon
        \,\left|\,P^\rho_{<1-\gamma/2}\right.\right)=0
    \qquad\forall\varepsilon>0.
$$
By continuity of $V$, this also implies
$$
    \lim_{t\to\infty}
    \mathbb{P}\left(V(\varphi_t(\rho,u_1))>\varepsilon
        \,\left|\,P^\rho_{<1-\gamma/2}\right.\right)=0
    \qquad\forall\varepsilon>0.
$$
As $V(\rho)\le 1$, we have
\begin{equation*}
\begin{split}
    \mathbb{E}\left(V(\varphi_t(\rho,u_1))\,\left|
        \,P^\rho_{<1-\gamma/2}\right.\right)
    \le &~
    \mathbb{P}\left(V(\varphi_t(\rho,u_1))>\varepsilon
        \,\left|\,P^\rho_{<1-\gamma/2}\right.\right) \\
    & ~~+\varepsilon\left[1-
    \mathbb{P}\left(V(\varphi_t(\rho,u_1))>\varepsilon
        \,\left|\,P^\rho_{<1-\gamma/2}\right.\right)\right].
\end{split}
\end{equation*}
Thus
$$
    \limsup_{t\to\infty}\,
    \mathbb{E}\left(V(\varphi_t(\rho,u_1))\,\left|
        \,P^\rho_{<1-\gamma/2}\right.\right)
    \le\varepsilon\qquad\forall\varepsilon>0
$$
which implies
$$
    \lim_{t\to\infty}
    \mathbb{E}\left(V(\varphi_t(\rho,u_1))\,\left|
        \,P^\rho_{<1-\gamma/2}\right.\right)
    =0.
$$
But we know by Theorem \ref{thm:localstab} that $V(\varphi_t(\rho,u_1))$
converges almost surely.  As $V$ is bounded, we obtain by dominated
convergence
$$
    \mathbb{E}\left(
    \lim_{t\to\infty}
        V(\varphi_t(\rho,u_1))\,\left|
        \,P^\rho_{<1-\gamma/2}\right.\right)
    =0
$$
from which the result follows immediately. \qquad
\end{proof}

\subsection*{Step 4}

It remains to combine the results of Steps 2 and 3 to prove existence,
uniqueness and global stability of the solution $\rho_t$. We will denote
by $u$ the control law of Theorem \ref{single:thm} and by
$\varphi_t(\rho,u)$ the associated solution.  Note that
$\varphi_t(\rho,u)$ is not a Markov process, as the control $u$ depends on
the past history of the solution.  We will construct $\varphi_t(\rho,u)$
by pasting together the strong Markov processes $\varphi_t(\rho,1)$ and
$\varphi_t(\rho,u_1)$ at the times where the control switches.

\begin{lemma}\label{final:lem}
    There is a unique solution $\varphi_t(\rho,u)$ for all 
    $t\in\mathbb{R}_+$.  Moreover, for almost every sample path of 
    $\varphi_t(\rho,u)$ there exists a time $T<\infty$ after which the 
    path never exits the set $\mathcal{S}_{<1-\gamma/2}$ and the active
    control law is $u_1$.
\end{lemma}

\begin{proof}
Fix the initial state $\rho$.  We begin by constructing a solution 
$\varphi_{t\wedge n}(\rho,u)$ up to (at most) an integer time 
$n\in\mathbb{N}$.  To this end, define the predictable stopping time 
$$
	\tau_1^n=\inf\{t\ge 0:\varphi_t(\rho,1)
		\in\mathcal{S}_{\le 1-\gamma}\}\wedge n.
$$
Then we can define $\rho_{\tau_1^n}=\varphi_{\tau_1^n}(\rho,1)$ 
and $\varphi_{t\wedge n}(\rho,u)=\varphi_t(\rho,1)$ for $t<\tau_1^n$.  In 
the following, we will need the two-parameter solution 
$\varphi_{s,t}(\rho,u')$ of the filtering equation under the simple 
control $u'$, given the initial state $\rho$ at time $s$.  Define
$$
	\sigma_1^n=\inf\{t\ge\tau_1^n:
	\varphi_{\tau_1^n,t}(\rho_{\tau_1^n},u_1)
		\in\mathcal{S}_{\ge 1-\gamma/2}\}\wedge n.
$$
We can extend our solution by
$$
	\varphi_{t\wedge n}(\rho,u)=\chi_{t<\tau_1^n}\varphi_{t}(\rho,1)+
		\chi_{\tau_1^n\le t<\sigma_1^n}
		\varphi_{\tau_1^n,t}(\rho_{\tau_1^n},u_1),
	\qquad t<\sigma_1^n
$$
where $\chi_A$ is the indicator function on the set $A$.  To extend the
solution further, we continue again with the control law $u=1$. 
Recursively, we define an entire sequence of predictable stopping times
$$
	\sigma_k^n=
	\inf\{t\ge\tau_k^n:
	\varphi_{\tau_k^n,t}(\rho_{\tau_k^n},u_1)
	\in\mathcal{S}_{\ge 1-\gamma/2}\}\wedge n,
$$
$$
	\tau_k^n=
	\inf\{t\ge\sigma_{k-1}^n:
	\varphi_{\sigma_{k-1}^n,t}(\rho_{\sigma_{k-1}^n},1)
	\in\mathcal{S}_{\le 1-\gamma}\}\wedge n,
$$
where
$$
	\rho_{\sigma_k^n}=
	\varphi_{\tau_k^n,\sigma_k^n}(\rho_{\tau_k^n},u_1),\qquad
	\rho_{\tau_k^n}=
	\varphi_{\sigma_{k-1}^n,\tau_k^n}(\rho_{\sigma_{k-1}^n},1).
$$
We can use these times to construct the solution
$$
	\varphi_{t\wedge n}(\rho,u)=\chi_{t<\tau_1^n}\varphi_t(\rho,1)
		+
		\sum_{k=1}^\infty
		\left[\chi_{\tau_k^n\le t<\sigma_k^n}
		\varphi_{\tau_k^n,t}(\rho_{\tau_k^n},u_1)
		+
		\chi_{\sigma_k^n\le t<\tau_{k+1}^n}
		\varphi_{\sigma_k^n,t}(\rho_{\sigma_k^n},1)
	\right]
$$
for all times $t<\Sigma^n=\lim_{k\to\infty}\sigma_k^n\le n$ (the limit 
exists, as $\sigma_k$ is a nondecreasing sequence of stopping times.)  
Moreover, the solution is a.s.\ unique, as the segments between each two 
stopping times are a.s.\ uniquely defined.

Now note that as anticipated by the notation, it is not difficult to
verify that $\varphi_{t\wedge (n+1)}(\rho,u)=\varphi_{t\wedge n}(\rho,u)$
a.s.\ for $t<\Sigma^n$, and moreover $\Sigma^n=\Sigma\wedge n$,
$\tau_k^n=\tau_k\wedge n$, $\sigma_k^n=\sigma_k\wedge n$ where
$\Sigma=\lim_{t\to\infty}\Sigma^n$ etc.  Hence we can let $n\to\infty$ to
obtain the unique solution $\varphi_t(\rho,u)$ defined up to the
accumulation time $\Sigma$, where $\tau_k$, $\sigma_k$ are the consecutive
times at which the control switches. It remains to prove that the solution
exists for all time, i.e.\ that $\Sigma=\infty$ a.s.  In particular, this
uniquely defines a c{\`a}dl{\`a}g control $u_t$, so that by uniqueness
$\varphi_t(\rho,u)$ must coincide with the solution of (\ref{eq:qfilt})  
with the control $u_t$.  Below we will prove that a.s., only finitely many
$\sigma_k$ are finite.  This is sufficient to prove not only existence,
but also the second statement of the Lemma.

To proceed, we use the fact that the strong Markov property holds on each 
segment between consecutive switching times $\tau_n\le t<\sigma_n$ or 
$\sigma_n\le t<\tau_{n+1}$.  Thus
\begin{equation*}
\begin{split}
    \mathbb{P}(\sigma_n<\infty&\mbox{ and }\tau_n<\infty)= \\
    &\int
        \chi_{\tau_n<\infty}(\tilde\omega)\,
        \mathbb{P}(\varphi_{t}(\rho_{\tau_n}(\tilde\omega),u_1)
        \mbox{ exits }\mathcal{S}_{<1-\gamma/2}\mbox{ in finite time})
    \,\mathbb{P}(d\tilde\omega)
\end{split}
\end{equation*}
which implies
\begin{equation*}
\begin{split}
    \mathbb{P}(\sigma_n<\infty&\,|\,\tau_n<\infty)= \\
    &\int
        \mathbb{P}(\varphi_{t}(\rho_{\tau_n}(\tilde\omega),u_1)
        \mbox{ exits }\mathcal{S}_{<1-\gamma/2}\mbox{ in finite time})
    \,\mathbb{P}(d\tilde\omega\,|\,\tau_n<\infty).
\end{split}
\end{equation*}
But $\rho_{\tau_n}\in\mathcal{S}_{\le 1-\gamma}$ on a set
$\Omega_{\tau_n}$ with $\mathbb{P}(\Omega_{\tau_n}\,|\,\tau_n<\infty)=1$.
Hence by Lemma \ref{sixth:lem}
$$
    \mathbb{P}(\sigma_n<\infty\,|\,\tau_n<\infty) \le 1-p.
$$
Through a similar argument, and using Lemma \ref{fifth:lem}, we obtain
$$
    \mathbb{P}(\tau_n<\infty\,|\,\sigma_{n-1}<\infty) = 1.
$$
But note that by construction
$$
    \mathbb{P}(\tau_n<\infty\,|\,\sigma_{n}<\infty)=
    \mathbb{P}(\sigma_{n-1}<\infty\,|\,\tau_n<\infty)=1.
$$
Hence we obtain
\begin{equation*}
\begin{split}
    \frac{\mathbb{P}(\sigma_n<\infty)}{\mathbb{P}(\sigma_{n-1}<\infty)}
    &=
    \frac{
        \mathbb{P}(\tau_n<\infty\,|\,\sigma_{n}<\infty)
        \mathbb{P}(\sigma_n<\infty)
    }{
         \mathbb{P}(\tau_n<\infty)
    }\,
    \frac{
        \mathbb{P}(\sigma_{n-1}<\infty\,|\,\tau_n<\infty)
        \mathbb{P}(\tau_n<\infty)
    }{
         \mathbb{P}(\sigma_{n-1}<\infty)
    } \\
    &=\mathbb{P}(\sigma_n<\infty\,|\,\tau_n<\infty)\,
    \mathbb{P}(\tau_n<\infty\,|\,\sigma_{n-1}<\infty)\le 1-p.
\end{split}
\end{equation*}
But $\mathbb{P}(\sigma_1<\infty)=
\mathbb{P}(\sigma_1<\infty\,|\,\tau_1<\infty)\le 1-p$ as $\tau_1<\infty$
a.s.  Hence
$$
    \mathbb{P}(\sigma_n<\infty)\le(1-p)^n
$$
and thus
$$
    \sum_{n=1}^\infty\mathbb{P}(\sigma_n<\infty)
    \le \sum_{n=1}^\infty(1-p)^n=\frac{1-p}{p}<\infty.
$$
By the Borel-Cantelli lemma, we conclude that
$$
    \mathbb{P}(\sigma_n<\infty\mbox{ for infinitely many }n)=0.
$$
Hence $\Sigma=\infty$ a.s.\ and for almost every sample path, there exists 
an integer $N<\infty$ such that $\sigma_n=\infty$ (and hence also 
$\tau_{n+1}=\infty$) for all $n\ge N$, and such that $\sigma_n<\infty$ 
(and hence also $\tau_{n+1}<\infty$) for all $n<N$, which implies the 
assertion.
\qquad
\end{proof}

Finally, we can now put together all the ingredients and complete the
proof of Theorem \ref{single:thm}.

{\em Proof of Theorem \ref{single:thm}}.
We must check three things: that the target state $\rho_f$ is (locally)
stable in probability; that almost all sample paths are attracted to
the target state as $t\to\infty$; and that this is also true in
expectation.  Existence and uniqueness of the solution follows from
Lemma \ref{final:lem}.

(i) To study local stability, we can restrict ourselves to the stopped
process
$$  \varphi_{t\wedge\tilde\tau}(\rho,u)=
    \varphi_{t\wedge\tilde\tau}(\rho,u_1),\quad
    \tilde\tau=\inf\{t:\varphi_t(\rho,u)\not\in\mathcal{S}_{<1-\gamma/2}\}.
$$
Denote by $\mathscr{\tilde A}$ the weak infinitesimal operator of
$\varphi_{t\wedge\tilde\tau}(\rho,u_1)$, and note that Prop.\
\ref{pro:stopopen} allows us to calculate $\mathscr{\tilde A}V$ from
(\ref{single:eq}) in the usual way.  In particular, we find
$\mathscr{\tilde A}V(\rho)=-u_1(\rho)^2\le 0$ for
$\rho\in\mathcal{S}_{<1-\gamma/2}$.  Hence we can apply Theorem
\ref{thm:localstab} with $Q_\lambda=\mathcal{S}_{<1-\gamma/2}$ to conclude
stability in probability.

(ii) From Lemmas \ref{seventh:lem} and \ref{final:lem}, it follows that
$\varphi_{t}(\rho,u)\to\rho_f$ a.s.\ as $t\to\infty$.

(iii) We have shown that
$$
    \mathbb{E}\left[
        \lim_{t\to\infty}V(\varphi_{t}(\rho,u))
    \right]=V(\rho_f)=0.
$$
But as $V$ is uniformly bounded, we obtain by dominated convergence
$$
    V\left(\lim_{t\to\infty}\mathbb{E}\varphi_{t}(\rho,u)\right)=
    \lim_{t\to\infty}\mathbb{E}\left[V(\varphi_{t}(\rho,u))\right]=0
$$
where we have used that $V$ is linear and continuous.  Hence
$\mathbb{E}\varphi_{t}(\rho,u)\to\rho_f$.
\qquad\endproof

\section{Two-qubit systems}
\label{multi:sec}

The methods employed in the previous section can be extended to other
quantum feedback control problems.  As an example, we treat the case of
two qubits in a symmetric dispersive interaction with an optical probe
field.  Qubits, i.e.\ two-level quantum systems (having a Hilbert space
of dimension two), and in particular correlated (entangled) states of
multiple such qubits, play an important role in quantum information
processing.  Here we investigate the stabilization of two such states in
the two-qubit system.

We begin by defining the Pauli matrices
$$
    \sigma_x=\left(
    \begin{array}{cc}
        0 & 1 \\ 1 & 0
    \end{array}
    \right),\qquad
    \sigma_y=\left(
    \begin{array}{cc}
        0 & -i \\ i & 0
    \end{array}
    \right),\qquad
    \sigma_z=\left(
    \begin{array}{cc}
        1 & 0 \\ 0 & -1
    \end{array}
    \right)
$$
and we define the basis $\psi_\uparrow=(1~0)^*$ and
$\psi_\downarrow=(0~1)^*$ in $\mathbb{C}^2$.  A system of two qubits
lives on the 4-dimensional space $\mathbb{C}^2\otimes\mathbb{C}^2$ with
the standard basis
\{$\psi_{\uparrow\uparrow}=\psi_\uparrow\otimes\psi_\uparrow$,
$\psi_{\uparrow\downarrow}=\psi_\uparrow\otimes\psi_\downarrow$,
$\psi_{\downarrow\uparrow}=\psi_\downarrow\otimes\psi_\uparrow$,
$\psi_{\downarrow\downarrow}=\psi_\downarrow\otimes\psi_\downarrow$\}.
We denote by $\sigma_{x,y,z}^1=\sigma_{x,y,z}\otimes\II$ and
$\sigma_{x,y,z}^2=\II\otimes\sigma_{x,y,z}$ the Pauli matrices on the
first and second qubit, respectively, and by
$F_{x,y,z}=\sigma_{x,y,z}^1+\sigma_{x,y,z}^2$ the (unnormalized)
collective angular momentum operators.

The quantum filtering equation for the two-qubit system is given by an
equation of the form (\ref{eq:qfilt}):
\begin{equation}\label{twoq:eq}
\begin{split}
    d\rho_t&=-iu_1(t)[\sigma_y^1,\rho_t]\,dt-iu_2(t)[\sigma_y^2,\rho_t]\,dt
    \\
    &\qquad-\tfrac{1}{2}[F_z,[F_z,\rho_t]]\,dt
    +\sqrt{\eta}\,(F_z\rho_t+\rho_t F_z-2\,\tr{F_z \rho_t}\rho_t)\,dW_t
\end{split}
\end{equation}
where $u_1$ and $u_2$ are two independent controls acting as local
magnetic fields in the $y$-direction on each of the qubits.
The main goal of this section is two stabilize this system around
two interesting target states,
$$
    \rho_s=\frac{1}{2}(\psi_{\uparrow\downarrow}+\psi_{\downarrow\uparrow})
        (\psi_{\uparrow\downarrow}+\psi_{\downarrow\uparrow})^*,
    \qquad
    \rho_a=\frac{1}{2}(\psi_{\uparrow\downarrow}-\psi_{\downarrow\uparrow})
        (\psi_{\uparrow\downarrow}-\psi_{\downarrow\uparrow})^*.
$$
Here $\rho_s$ is a symmetric and $\rho_a$ is an antisymmetric qubit state.

\begin{theorem}\label{main2:thm}
Consider the following control law:
\begin{enumerate}
    \item $u_1(t)=1-\tr{i[\sigma_y^1,\rho_t]\rho_a},~
    u_2(t)=1-\tr{i[\sigma_y^2,\rho_t]\rho_a}$
     if $\tr{\rho \rho_a}\ge\gamma$;
    \item $u_1(t)=1,~u_2(t)=0$ if $\tr{\rho\rho_a}\le\gamma/2$;
    \item If $\rho_t\in\mathcal{B}_a=\{\rho:\gamma/2<\tr{\rho
    \rho_a}<\gamma\}$,
    then take $u_1(t)=1-\tr{i[\sigma_y^1,\rho_t]\rho_a}$,
    $u_2(t)=1-\tr{i[\sigma_y^2,\rho_t]\rho_a}$ if $\rho_t$
    last entered the set $\mathcal{B}_a$ through the boundary
    $\tr{\rho\rho_a}=\gamma$, and $u_1(t)=1,~u_2(t)=0$ otherwise.
\end{enumerate}
Then $\exists\gamma>0$ s.t.\ {\rm (\ref{twoq:eq})} is globally stable
around $\rho_a$ and $\mathbb{E}\rho_t\to\rho_a$ as $t\to\infty$.
Similarly,
\begin{enumerate}
    \item $u_1(t)=1-\tr{i[\sigma_y^1,\rho_t]\rho_s},~
    u_2(t)=-1-\tr{i[\sigma_y^2,\rho_t]\rho_s}$
     if $\tr{\rho \rho_s}\ge\gamma$;
    \item $u_1(t)=1,~u_2(t)=0$ if $\tr{\rho\rho_s}\le\gamma/2$;
    \item If $\rho_t\in\mathcal{B}_s=\{\rho:\gamma/2<\tr{\rho
    \rho_s}<\gamma\}$,
    then take $u_1(t)=1-\tr{i[\sigma_y^1,\rho_t]\rho_s}$,
    $u_2(t)=-1-\tr{i[\sigma_y^2,\rho_t]\rho_s}$ if $\rho_t$
    last entered the set $\mathcal{B}_s$ through the boundary
    $\tr{\rho\rho_s}=\gamma$, and $u_1(t)=1,~u_2(t)=0$ otherwise.
\end{enumerate}
stabilizes the system around the symmetric state $\rho_s$.
\end{theorem}

We will prove the result for the antisymmetric case; the proof
for the symmetric case may be done exactly in the same manner.
We proceed in the same way as in the proof of Theorem~\ref{single:thm}.

\subsection*{Step 1}

The proof of Lemma \ref{first:lem} carries over directly to the two
qubit case.  The proof of Lemma \ref{add:lem} also carries over after
minor modifications; in particular, in the two qubit case we can
explicitly compute that
$$
    A=-i\sigma_y^1-F_z^2+2F_z=
     \begin{pmatrix}
        0 & -1  & 0 &0 \\
        1 & 0 &0& 0\\
        0 & 0& 0& -1\\
        0 & 0& 1& -8\\
      \end{pmatrix}
$$
admits the diagonlization $A=PDP^{-1}$ with
$$
    P=
     \begin{pmatrix}
        1 & 1  & 0 &0 \\
        -i & i &0& 0\\
        0 & 0& 1& 1\\
        0 & 0& .1270& 7.8730\\
      \end{pmatrix},\qquad
     D=\begin{pmatrix}
        i & 0  & 0 &0 \\
        0 & -i &0& 0\\
        0 & 0& -.1270& 0\\
        0 & 0& 0& -7.8730\\
      \end{pmatrix}.
$$
Hence the matrix $A$ has a nondegenerate spectrum and moreover
$$
    \tilde v_a=\tfrac{1}{\sqrt 2}\,
        P^*(\psi_{\uparrow\downarrow}-\psi_{\downarrow\uparrow})
        =\tfrac{1}{\sqrt 2}\,(i ~ -i ~ -1 ~ -1)^*
$$
has only nonzero entries.  The remainder of the proof is identical to
that of Lemma~\ref{first:lem}.

\subsection*{Step 2}

The proofs of Lemmas~\ref{fourth:lem} and \ref{fifth:lem} carry over
directly.

\subsection*{Step 3}

The proofs of Lemmas \ref{sixth:lem} and \ref{seventh:lem} carry over
directly.  The following replaces Lemma \ref{seventh2:lem}.  We denote by
$U_1(\rho)=1-\tr{i[\sigma_y^1,\rho]\rho_a}$,
$U_2(\rho)=1-\tr{i[\sigma_y^2,\rho]\rho_a}$ and by
$\varphi_t(\rho,U_1,U_2)$ the associated solution of (\ref{twoq:eq}).

\begin{lemma}\label{qu:lem}
The sample paths of $\varphi_t(\rho,U_1,U_2)$ that never exit the set
$\SSS_{<1-\gamma/2}$ converge in probability to $\rho_a$ as
$t\to\infty$.
\end{lemma}

\begin{proof}
Consider the Lyapunov function
$$
    \VV(\rho)=1-\tr{\rho\rho_a}^2.
$$
It is easily verified that $\VV(\rho)\geq 0$ for all $\rho\in\SSS$ and
that $\VV(\rho)=0$ iff $\rho=\rho_a$. A straightforward computation gives
\begin{equation*}
    \mathscr{A}\VV(\rho)=
    -2\left[
        (U_1(\rho)-1)^2+(U_2(\rho)-1)^2
    \right]\tr{\rho\rho_a}
    -4\eta\,\tr{\rho F_z}^2\,\tr{\rho\rho_a}^2
    \le 0
\end{equation*}
where $\mathscr{A}$ is the weak infinitesimal operator associated
to $\varphi_t(\rho,U_1,U_2)$ (here we have used $[F_y,\rho_a]=0$
in calculating this expression). Now note that all the conditions
of Theorem \ref{thm:lasalle} are satisfied by virtue of Prop.\
\ref{pro:feller} and \ref{pro:uniformstoch}.  Hence
$\varphi_t(\rho,U_1,U_2)$ converges in probability to the largest
invariant set contained in
$\mathcal{C}=\{\rho\in\mathcal{S}:\mathscr{A}\VV(\rho)=0\}$.

In order to satisfy the condition $\mathscr{A}\VV(\rho)=0$ we must have
at least
$$
    \text{either}\quad \tr{\rho\rho_a}=0 \qquad \text{or} \quad
    \tr{\rho F_z}=0.
$$
Let us investigate the largest invariant set contained in
$\mathcal{C}'=\{\rho\in\SSS:\tr{\rho F_z}=0\}$.  Clearly this
invariant set can only contain $\rho\in\mathcal{C}'$ for which
$\tr{\varphi_t(\rho,U_1,U_2)F_z}$ is constant.  Using It\^o's rule we
obtain
$$
    d\,\tr{\rho_tF_z}=-\sum_{j=1}^2
        U_j(\rho_t)\,\tr{i[\sigma_y^j,\rho_t]F_z}\,dt
        +2\sqrt{\eta}\,(\tr{F_z^2\rho_t}-\tr{F_z\rho_t}^2)\,dW_t.
$$
Hence in order for $\tr{\varphi_t(\rho,U_1,U_2)F_z}$ to be constant, we
must at least have
$$
    \tr{F_z^2\rho}-\tr{F_z\rho}^2=0
$$
which implies that $\rho$ must be an eigenstate of $F_z$.
The latter can only take one of the following forms: either
$\rho=\psi_{\uparrow\uparrow}\psi_{\uparrow\uparrow}^*$ or
$\rho=\psi_{\downarrow\downarrow}\psi_{\downarrow\downarrow}^*$, or
$\rho$ is any state of the form
\begin{equation}\label{mixedset:eq}
    \rho=
    \alpha\psi_{\uparrow\downarrow}\psi_{\uparrow\downarrow}^*+
    \beta\psi_{\uparrow\downarrow}\psi_{\downarrow\uparrow}^*+
    \beta^*\psi_{\downarrow\uparrow}\psi_{\uparrow\downarrow}^*+
    (1-\alpha)\psi_{\downarrow\uparrow}\psi_{\downarrow\uparrow}^*.
\end{equation}
Let us investigate in particular the latter case.  Note that any density
matrix of the form (\ref{mixedset:eq}) satisfies $F_z\rho=\rho F_z=0$.
Suppose that (\ref{twoq:eq}) with $u_1=U_1$, $u_2=U_2$ leaves the set
(\ref{mixedset:eq}) invariant; then the solution at time $t$ of
\begin{equation}\label{eq:deteffeq}
    \frac{d}{dt}\rho_t=-i[F_y,\rho_t]
\end{equation}
must coincide with $\varphi_t(\rho,U_1,U_2)$ when $\rho$ is of the
form (\ref{mixedset:eq}), and in particular (\ref{eq:deteffeq})
must leave the set (\ref{mixedset:eq}) invariant (here we have
used that $U_1(\rho)=U_2(\rho)=1$ for $\rho$ of the form
(\ref{mixedset:eq})). We claim that this is only the case if
$\rho=\rho_a$, which implies that of all states of the form
(\ref{mixedset:eq}) only $\rho_a$ is in fact invariant.  To see
this, note that by Lemma \ref{l:hullc} we can write any $\rho$ of
the form (\ref{mixedset:eq}) as a convex combination
$\sum_i\lambda_i\psi^i\psi^{i*}$ of unit vectors $\psi^i\in {\rm
span}\{\psi_{\uparrow\downarrow},\psi_{\downarrow\uparrow}\}$.
Thus the solution of (\ref{eq:deteffeq}) at time $t$ is given by
$\sum_i\lambda_i\psi_t^i\psi_t^{i*}$ with
\begin{equation*}
    \frac{d}{dt}\psi_t^i=-iF_y\psi_t^i,\qquad\psi_0^i=\psi^i.
\end{equation*}
But $F_y\psi^i\not\in{\rm span}\{\psi_{\uparrow\downarrow},
\psi_{\downarrow\uparrow}\}$ unless $\psi^i\propto
\psi_{\uparrow\downarrow}-\psi_{\downarrow\uparrow}$, which
implies the assertion.

From the discussion above it is evident that the largest invariant
set contained in $\mathcal{C}$ must be contained inside the set
$\{\rho_a\}\cup\mathcal{S}_1$.  But then the paths that never exit
$\mathcal{S}_{<1-\gamma/2}$ must converge in probability to
$\rho_a$. Thus the Lemma is proved.\qquad
\end{proof}

\subsection*{Step 4}

The remainder of the proof of Theorem~\ref{main2:thm} carries over
directly.

\section*{Acknowledgments}

The authors thank Hideo Mabuchi and Houman Owhadi for helpful
discussions.

\bibliography{./rhn}

\begin{thebibliography}{10}

\bibitem{arnold}
L.\ Arnold.
\newblock {\em Stochastic Differential Equations: Theory and Applications}.
\newblock Wiley, 1974.

\bibitem{belavkin}
V.~P. Belavkin.
\newblock Quantum stochastic calculus and quantum nonlinear filtering.
\newblock {\em J. Multivariate Anal.}, 42:171--201, 1992.

\bibitem{Bensoussan1992}
A.~Bensoussan.
\newblock {\em Stochastic Control of Partially Observable Systems}.
\newblock Cambridge University Press, 1992.

\bibitem{bouten}
L.~Bouten, M.~Gu\c{t}\u{a}, and H.~Maassen.
\newblock Stochastic {S}chr{\"o}dinger equations.
\newblock {\em J. Phys. A: Math. Gen.}, 37:3189--3209, 2004.

\bibitem{BvHJ-05}
L.~Bouten, R.~Van Handel, and M.~R. James.
\newblock An introduction to quantum filtering.
\newblock {\em In preparation; see {\tt http://arxiv.org/abs/math-ph/0508006}},
  2005.

\bibitem{dynkin-book1}
E.B. Dynkin.
\newblock {\em Markov Processes}, volume~I.
\newblock Springer-Verlag, 1965.

\bibitem{Geremia-science}
J.~M. Geremia, J.~K. Stockton, and H.~Mabuchi.
\newblock Real-time quantum feedback control of atomic spin-squeezing.
\newblock {\em Science}, 304:270--273, 2004.

\bibitem{gikhman}
I.~I. Gikhman and A.~V. Skorokhod.
\newblock {\em Introduction to the theory of random processes}.
\newblock Dover, 1996.

\bibitem{vanhandel-05b}
R.~Van Handel and H.~Mabuchi.
\newblock Quantum projection filter for a highly nonlinear model in cavity
  {QED}.
\newblock {\em J. Opt. B: Quantum Semiclass. Opt.}, 7:S226--S236, 2005.

\bibitem{vanhandel-05}
R.~Van Handel, J.~K. Stockton, and H.~Mabuchi.
\newblock Feedback control of quantum state reduction.
\newblock {\em IEEE Trans. Automat. Control}, 50:768--780, 2005.

\bibitem{vanhandel-review}
R.~Van Handel, J.~K. Stockton, and H.~Mabuchi.
\newblock Modelling and feedback control design for quantum state preparation.
\newblock {\em J. Opt. B: Quantum Semiclass. Opt.}, 7:S179--S197, 2005.

\bibitem{hasminskii}
R.~Z. Has'minski\u{\i}.
\newblock {\em Stochastic stability of differential equations}.
\newblock Sijthoff \& Noordhoff, 1980.

\bibitem{kunita-supp}
H.~Kunita.
\newblock Supports of diffusion processes and controllability problems.
\newblock In {\em Proc. Intern. Symp. SDE, Kyoto, 1976}, pages 163--185, 1978.

\bibitem{kunita-flow}
H.~Kunita.
\newblock {\em Stochastic flows and stochastic differential equations}.
\newblock Cambridge, 1990.

\bibitem{kushner-67}
H.~J. Kushner.
\newblock {\em Stochastic Stability and Control}.
\newblock Academic Press, 1967.

\bibitem{kushner-68}
H.~J. Kushner.
\newblock The concept of invariant set for stochastic dynamical systems and
  applications to stochastic stability.
\newblock In H.~F. Karreman, editor, {\em Stochastic Optimization and Control},
  pages 47--57. Wiley, 1968.

\bibitem{kushner-72}
H.~J. Kushner.
\newblock Stochastic stability.
\newblock In R.F. Curtain, editor, {\em Stability of Stochastic Dynamical
  systems}, volume 294 of {\em Lecture Notes in Mathematics}, pages 97--123.
  Springer-Verlag, 1972.

\bibitem{maassen-qprob}
H.~Maassen.
\newblock Quantum probability applied to the damped harmonic oscillator.
\newblock In S.~Attal and J.~M. Lindsay, editors, {\em Quantum Probability
  Communications {XII}}, pages 23--58. World Scientific, 2003.

\bibitem{merzbacher}
E.~Merzbacher.
\newblock {\em Quantum mechanics}.
\newblock Wiley, third edition, 1998.

\bibitem{MvHMM-05}
M.~Mirrahimi, R.~Van Handel, A.~E. Miller, and H.~Mabuchi, 2005.
\newblock In preparation.

\bibitem{mirrahimi-et-al2-04}
M.~Mirrahimi, P.~Rouchon, and G.~Turinici.
\newblock Lyapunov control of bilinear {S}chr{\"o}dinger equations.
\newblock {\em Automatica}, 2005.
\newblock At press.

\bibitem{oksendal}
B.~{\O}ksendal.
\newblock {\em Stochastic Differential Equations}.
\newblock Springer, fifth edition, 1998.

\bibitem{protter}
P.~E. Protter.
\newblock {\em Stochastic Integration and Differential Equations}.
\newblock {Springer}, second edition, 2004.

\bibitem{stroock-varadhan}
D.~W. Stroock and S.~R. Varadhan.
\newblock On the support of diffusion processes with applications to the strong
  maximum principle.
\newblock In {\em Proc. 6th Berkely Sympos. Math. Statist prob.}, volume III,
  pages 333--368, 1972.

\end{thebibliography}

\end{document}